\documentclass[fleqn,usenatbib]{mnras}

\usepackage{newtxtext,newtxmath}

\usepackage[T1]{fontenc}

\DeclareRobustCommand{\VAN}[3]{#2}
\let\VANthebibliography\thebibliography
\def\thebibliography{\DeclareRobustCommand{\VAN}[3]{##3}\VANthebibliography}

\usepackage{graphicx}	
\usepackage{amsmath}		
\usepackage{comment}
\usepackage[normalem]{ulem}

\title[Energetic PUIs downstream of TS as revealed by {\it IBEX-Hi}]{Energetic pickup proton population downstream of the termination shock as revealed by the {\it IBEX-Hi} data}

\author[I. I. Baliukin et al.]{
I. I. Baliukin,$^{1,2,3}$\thanks{E-mail: igor.baliukin@gmail.com}
V. V. Izmodenov,$^{1,2,4}$
and D. B. Alexashov$^{1,4}$
\\
$^{1}$Space Research Institute of Russian Academy of Sciences, Profsoyuznaya Str. 84/32, Moscow, 117997, Russia\\
$^{2}$Lomonosov Moscow State University, Moscow Center for Fundamental and Applied Mathematics, GSP-1, Leninskie Gory, Moscow, 119991, Russia\\
$^{3}$HSE University, Moscow, Russia\\
$^{4}$Institute for Problems in Mechanics, Vernadskogo 101-1, Moscow, 119526, Russia
}

\date{Accepted XXX. Received YYY; in original form ZZZ}

\pubyear{2021}

\begin{document}
\label{firstpage}
\pagerange{\pageref{firstpage}--\pageref{lastpage}}
\maketitle

\begin{abstract}
The pickup protons originate as a result of the ionization of hydrogen atoms in the supersonic solar wind, forming the suprathermal component of protons in the heliosphere. While picked by the heliospheric magnetic field and convected into the heliosheath, the pickup protons may suffer the stochastic acceleration by the solar wind turbulence in the region from the Sun up to the heliospheric termination shock, where they can also experience the shock-drift acceleration or the reflection from the cross-shock potential. These processes create a high-energy <<tail>> in the pickup ion energy distribution. The properties of this energetic pickup proton population are still not well-defined, despite they are vital for the models to simulate energetic neutral atom fluxes. We have considered two scenarios for pickup proton velocity distribution downstream of the heliospheric termination shock (filled shell with energetic power-law <<tail>> and bi-Maxwellian). Based on the numerical kinetic model and observations of the energetic neutral atom fluxes from the inner heliosheath by the {\it IBEX-Hi} instrument, we have characterized the pickup proton distribution and provided estimations on the properties of the energetic pickup proton population downstream of the termination shock.

\end{abstract}

\begin{keywords}
ISM: atoms --- ISM: magnetic fields --- Sun: heliosphere
\end{keywords}



\section{Introduction} \label{sec:intro}

The interaction of the supersonic solar wind (SW) with the ionized component of the local interstellar medium (LISM) shapes the heliospheric interface. The tangential discontinuity, the heliopause (HP), divides the solar wind and interstellar plasmas, and the heliospheric termination shock (TS) is the boundary at which the solar wind is slowed down. 
The TS crossings from Voyager 1 \& 2 occurred in 2004 and 2007, respectively, at distances of $\sim$94 AU \citep{decker2005, stone2005} and $\sim$84 au \citep{decker2008, stone2008}. The Voyager 1 \& 2 crossings of the HP occurred in 2012 \citep{krimigis2013, stone2013, burlaga2013, gurnett2013} and 2018 \citep{krimigis2019, stone2019, richardson2019, gurnett2019, burlaga2019} at $\sim$122 au and $\sim$119 au, respectively.
The region of perturbed solar wind plasma between the termination shock and the heliopause is called the inner heliosheath (IHS).

In addition to the thermal component of solar wind protons, there is the suprathermal component -- the pickup protons, which originate as a result of the ionization (mainly due to the process of charge exchange) of hydrogen atoms. The H atoms penetrate the heliosphere due to the Sun--LISM relative motion and their large mean free path for charge exchange that is comparable with the characteristic size of the heliosphere \citep[e.g.][]{izmod2001}. The pickup ions (PUIs) that have been created in the supersonic solar wind are picked by the heliospheric magnetic field and convected into the heliosheath. They may suffer the stochastic acceleration by the solar wind turbulence in the region from the Sun up to the termination shock \citep{fisk1976, isenberg1987, bogdan1991, chalov1995, chalov1997, fichtner1996, leroux1998} or experience acceleration at propagating interplanetary shocks \citep{giacalone1997}, which creates high-energy <<tail>> in the pickup ion energy distribution in the supersonic SW. However, an energetic <<tail>> can be formed without the stochastic acceleration through the ionization of energetic neutral atoms (ENAs), which, in turn, are produced in the process of charge exchange of protons and hydrogen atoms in the heliosheath \citep{chalov_fahr2003,gruntman2004}. Therefore, when the pickup protons reach the termination shock, they already have a pronounced high-velocity <<tail>>, which is required for the pickup protons to enter the regime of the drift acceleration at the TS. 
Important to mention that the observations made by Solar Wind Around Pluto (SWAP) instrument on New Horizons from 22 to 38 au showed an increase with distance of the temperature and thermal pressure of hydrogen pickup ions, suggesting some form of additional heating, which may be induced by compression of plasma and driven by the faster solar wind parcels overtaking slower parcels \citep{mccomas2017}.

The Voyager 2 observations at the crossing of the TS showed that the downstream thermal protons still move with supersonic speed \citep{richardson2008}, which is due to the multifluid nature of the solar wind  \citep[the pickup ions absorb most of the upstream kinetic solar wind energy; see, e.g., ][]{chalov_fahr2010}. 
The interaction of PUIs with the heliospheric termination shock is complex, and it is forced by the physical processes that are still under discussion. On arrival at the TS, some portion of pickup protons (with high enough velocities) can experience the diffusive shock-drift acceleration, which is most effective at the flanks of the heliosphere \citep{chalov2005,mccomas2006,giacalone2010, chalov2012, chalov2015}. The pickup ions with small velocities, in turn, can suffer the reflection from the cross-shock potential \citep[so-called <<shock-surfing>> mechanism; see][]{lee1996, zank1996, zank2010, burrows2010}.

For the reason that the pickup protons are the seed population for the energetic neutral atoms, the observations of ENA fluxes carry information about the distribution of PUIs in the region of their creation. The fluxes of energetic neutral atoms have been measured from the Earth's orbit by the {\it IBEX-Hi} instrument \citep[0.3 -- 6 keV;][]{funsten2009} onboard the {\it Interstellar Boundary Explorer (IBEX)} for more than a decade \citep{mccomas2020}. These observations have revealed two distinct populations of ENA fluxes \citep{mccomas2009}: the so-called <<ribbon>>, which is emitted from a narrow circular part of the sky and formed by the secondary charge exchange in the outer heliosheath \citep[see, e.g., ][]{mccomas2009,chalov2010}, and a globally distributed flux (GDF) populated by the energetic neutral atoms that originate in the inner heliosheath \citep{schwadron2011}.







In this work, we do not investigate the physical processes in the heliosphere that produce high-energy <<tail>> in the energy distribution of pickup protons but set ourselves a goal to derive properties of the energetic population of pickup protons in the inner heliosheath. 
We determine the parameters of this population by comparison of the {\it IBEX-Hi} GDF data with the results of parametric numerical modelling. We use \citet{baliukin2020} model, which treats PUIs kinetically and employs global plasma and neutral distributions from \citet{izmod2020} heliospheric model, and in this study, two scenarios for PUI distribution downstream the TS are considered.

Section \ref{sec:model} describes the model of pickup proton distribution in the heliosphere that accounts for the energetic population of PUIs. Section \ref{sec:fitting} presents the algorithm for fitting the {\it IBEX-Hi} data using the developed model. In Section \ref{sec:results}, the results of the work with the estimations on the parameters of the energetic pickup proton population are presented. Section \ref{sec:conclusions} provides a summary of the work along with a discussion.

\section{Model} \label{sec:model}
In this section, we describe the kinetic model of PUI distribution in the heliosphere that accounts for the additional energetic population of pickup ions. Using the method described in \citet{baliukin2020}, the PUI velocity distribution function, which is assumed to be isotropic in the SW reference frame, can be calculated everywhere downstream of the TS. The interaction of pickup ions with the TS is taken into account by means of the jump condition that is based on (i) the Liouville’s theorem, (ii) the conservation of the magnetic moment,
and (iii) the assumption of the weak scattering \citep[for details, see][]{fahr_siewert2011,fahr_siewert2013}. The distribution downstream of the termination shock, in this case, has the form of the so-called filled shell \citep{vasyliunas1976, zank2010} that is compressed at the TS.

The formal solution of the kinetic equation for isotropic velocity distribution function of pickup protons \citep[see Equation 3 in][]{baliukin2020} in the IHS with the boundary condition at the TS can be written as:
\begin{equation}
\begin{split}
f_{\rm pui}^{*}(t, \mathbf{r}, w) 
&=  \int^t_{t_{\rm TS}} S_+(\tau, \mathbf{r}(\tau), w(\tau)) S_{\rm p,pui}(\tau, t) {\rm d} \tau \\
&+ f_{\rm pui,d}^{*}(t_{\rm TS}, \mathbf{r}(t_{\rm TS}), w(t_{\rm TS})) S_{\rm p,pui}(t_{\rm TS}, t).
\end{split}
\label{eq:solution_refl}
\end{equation}
where $f_{\rm pui}^{*}$ is the value of PUI velocity distribution function at particular moment $t$ and point $\mathbf{r}$ in the IHS, $f_{\rm pui,d}^{*}$ is the value of PUI velocity distribution function downstream of the TS, $w = \lvert \mathbf{v} - \mathbf{V}_{\rm p} \rvert$ is pickup velocity in the SW reference frame ($\mathbf{v}$ and $\mathbf{V}_{\rm p}$ are pickup proton velocity and
bulk velocity of the plasma in the Sun inertial reference frame), $t_{\rm TS}$ and $\mathbf{r}_{\rm TS} = \mathbf{r}(t_{\rm TS})$ are the moment and position of TS crossing, $S_{\rm +}$ is the source term responsible for the production of PUIs, and $S_{\rm p,pui}$ describes the loss of pickup ions due to neutralization (charge exchange with H atoms). The superscript <<*>> hereafter indicates the SW reference frame. In order to not overcharge the paper with expressions for the source and loss terms, we refer to \citet[][see its Equations 4 and 11]{baliukin2020}. 

The integration in Equation (\ref{eq:solution_refl}) is performed along the PUI trajectory defined by
\begin{equation}
\frac{{\rm d} \mathbf{r}}{{\rm d} t} = \mathbf{V}_{\rm p},
\end{equation}
with the pickup proton velocity (in the SW reference frame) change according to
\begin{equation}
\frac{{\rm d} w}{{\rm d} t} = -\frac{w}{3} {\rm div}(\mathbf{V}_{\rm p}).
\end{equation}
Therefore, our model takes into account the adiabatic heating due to the compression of decelerated plasma when it moves from the termination shock to the heliopause, as opposed to the models by \citet{zirnstein2017} and \citet{kornbleuth2020} where this process was neglected. Let us also explicitly emphasize that the solution (\ref{eq:solution_refl}) implies the neglect of the spatial diffusion, which is reasonable for energies of $\sim$few keVs \citep[see][]{rucinski1993, chalov1997}, and also the velocity diffusion. However, the authors admit that the stochastic acceleration may also operate in the inner heliosheath. 



The first term of Equation (\ref{eq:solution_refl}) describes the origin of PUIs that experienced charge exchange in the IHS and injected at energies $\sim$0.1 keV, while the second term represents pickup ions that originated in the supersonic solar wind, convected into the heliosheath, and undergo extinction in this region.


To carry out modelling of the pickup proton distribution in the heliosphere, the distributions of hydrogen atoms and plasma in the heliosphere should be predetermined. In our study, the plasma and neutral distributions were obtained in the frame of the global heliospheric time-dependent model by \citet{izmod2020}.
Let us note that, as further advancement of the work by \citet{baliukin2020}, these distributions were utilized on the original \citet{izmod2015} numerical grid without spatial re-interpolation to the spherical grid.

Using the known distribution of PUIs, the directional differential flux, which is the line-of-sight (LOS) integral, of energetic neutral atoms can be simulated \citep[like it was performed in][see its Section 2 and Appendix A]{baliukin2020}, since the ENAs are produced in the process of charge exchange of PUIs and SW protons with the H atoms.
The integration of the ENA sources for the specified LOS is performed from the TS to the HP directly (distances to the TS and HP are calculated using the time-dependent model), and for the directions close to the downwind, where the HP distance is large, we limit the integration distance by 1500 au from the Sun. The TS distances calculated in the frame of the stationary version of the \citet{izmod2020} model, in the upwind, Voyager 1 \& 2, north and south ecliptic pole directions are 76, 85, 86, 114, and 112 au, while the corresponding HP distances are 117, 127, 130, 219, and 205 au, respectively. The stationary version of the heliospheric model utilizes the averaged solar cycle (from 1995 to 2017) boundary conditions at 1 au for the solar wind parameters \citep[see, Appendix A in][]{izmod2020}. The comparison of the TS and HP distances measured by Voyager 1 \& 2 with the time-dependent model results, can be found in \citet[][see its Fig. 2]{izmod2020}. The estimations for the HP distances toward the ecliptic poles obtained by \citet{reisenfeld2021} based on the IBEX ENA data collected over a complete solar cycle (from 2009 through 2019) are $\sim$160 -- 180 au, which are somewhat smaller compared to the model distances.

The main limitation of the model by \citet{baliukin2020} is the absence of processes that produce high-energy <<tail>> in PUI velocity distribution.
To see how the additional energetic population of pickup protons affects the modeled ENA fluxes, we consider two cases with modification of the PUI distribution downstream of the TS: (i) the filled shell distribution with a power-law <<tail>> representing the additional energetic PUIs, and (ii) the bi-Maxwellian distribution.

\subsection{<<Power-law tail>> scenario} \label{sec:powertail}
To model the energetic pickup proton population, the approach proposed by \citet[][see its Section 6]{baliukin2020} is utilized. Immediately downstream of the TS the distribution function is assumed to be the sum of the filled shell distribution $f_{\rm sh}^{*}$ (for velocities smaller than the cutoff velocity $w_{\rm c}$) and the energetic tail with a power-law distribution $f_{\rm tail}^{*}(w) \propto w^{-\eta}$ (for velocities higher than $w_{\rm c}$).
To be more specific, 
\begin{equation}
f_{\rm pui, d}^{*} =(1 - \xi) f_{\rm sh}^{*} + \xi f_{\rm tail}^{*},
\label{eq:bound_cond_tail}
\end{equation}
\begin{equation}
    f_{\rm tail}^{*}(t_{\rm TS}, \mathbf{r}_{\rm TS}, w) =
    \begin{cases}
      n_{\rm pui}(t_{\rm TS}, \mathbf{r}_{\rm TS}) \frac{\eta - 3}{4 \pi w_{\rm c}^3} \left(\frac{w}{w_{\rm c}}\right)^{-\eta}, & w \geq w_{\rm c} \\
      0, & w < w_{\rm c}
    \end{cases}
\label{eq:ftail}
\end{equation}
where $\eta$ is the index that specifies the inclination of the <<tail>> ($\eta > 3$, otherwise the number density of the PUIs in the <<tail>> diverges), $\xi$ is the density fraction of PUIs of the <<tail>> distribution, and the local PUI number density downstream of the TS is given by
\begin{equation}
n_{\rm pui}(t_{\rm TS}, \mathbf{r}_{\rm TS}) = 4 \pi \int f_{\rm sh}^{*} (t_{\rm TS}, \mathbf{r}_{\rm TS}, w) w^2 {\rm d} w.
\end{equation}

In the IHS, the solution (\ref{eq:solution_refl}) with the boundary condition downstream of the TS (\ref{eq:bound_cond_tail}) is employed. The second term of this condition is associated with the additional energetic population of pickup ions that have been accelerated. 
The proposed method of modification of the PUI velocity distribution function downstream of the TS conserves the total number density but not the total pressure. With the <<tail>> distribution assumed, some energy is <<injected>> into the system, which can be provided by the acceleration processes (such as the interaction of pickup ions with fluctuating heliospheric magnetic field) that are not taken into account in our modelling directly.



\subsection{<<Bi-Maxwellian>> scenario} \label{sec:bimaxwellian}
Let us consider the other scenario for the velocity distribution function downstream of the TS. 
Right after the TS, we assume the existence of two distinct populations of pickup protons -- transmitted and reflected. We additionally assume that these populations are co-moving with the plasma bulk velocity and their velocity distribution functions are isotropic. Therefore, according to the mass and thermal pressure balances
\begin{equation}
n_{\rm pui} = n_{\rm pui}^{\rm tr} + n_{\rm pui}^{\rm ref},\\
n_{\rm pui}T_{\rm pui} = n_{\rm pui}^{\rm tr} T_{\rm pui}^{\rm tr} + n_{\rm pui}^{\rm ref} T_{\rm pui}^{\rm ref},
\label{eq:pui_partition}
\end{equation}
where superscripts <<tr>> and <<ref>> denote transmitted and reflected PUI populations, respectively. We also introduce two additional parameters 
\begin{equation}
\alpha = \frac{n_{\rm pui}^{\rm ref}}{n_{\rm pui}},\: \beta = \frac{n_{\rm pui}^{\rm ref} T_{\rm pui}^{ \rm ref}}{n_{\rm pui} T_{\rm pui}},
\label{eq:pui_fractions}
\end{equation}
which are the density and thermal pressure fractions of reflected PUIs downstream of the TS. 
Using the introduced parameters (\ref{eq:pui_fractions}) and equations (\ref{eq:pui_partition}), the moments of reflected and transmitted particles can be expressed in terms of the moments of the total pickup ion population:
\begin{equation}
n_{\rm pui}^{\rm ref} = \alpha n_{\rm pui},\: T_{\rm pui}^{\rm ref} = \frac{\beta}{\alpha} T_{\rm pui},
\end{equation}
\begin{equation}
n_{\rm pui}^{\rm tr} = (1 - \alpha) n_{\rm pui},\: T_{\rm pui}^{\rm tr} = \frac{1 - \beta}{1 - \alpha} T_{\rm pui}.
\end{equation}
We set the boundary condition for the velocity distribution function immediately after the TS as the sum of two isotropic Maxwellians in the solar wind rest frame:
\begin{equation}
\begin{split}
f_{\rm pui,d}^{*}(t_{\rm TS}, \mathbf{r}_{\rm TS}, w) 
&= \frac{n_{\rm pui}^{\rm tr}}{(c_{\rm tr} \sqrt{\pi})^3} \exp \left(-\frac{w^2}{c_{\rm tr}^2}\right) \\
&+ \frac{n_{\rm pui}^{\rm ref}}{(c_{\rm ref} \sqrt{\pi})^3} \exp \left(-\frac{w^2}{c_{\rm ref}^2}\right),
\end{split}
\label{eq:bound_cond_mxwl}
\end{equation}
where $c_{\rm i} = \sqrt{2 k_{\rm B} T_{\rm pui}^{\rm i} / m_{\rm p}},\: {\rm i} \in \{\rm tr, ref\}$ defines the transmitted and reflected thermal velocities. We also note, that in the case of $\alpha = \beta$, the bi-Maxwellian boundary condition (\ref{eq:bound_cond_mxwl}) transforms into a single Maxwellian distribution function with number density $n_{\rm pui}$ and temperature $T_{\rm pui}$. Even though the assumption of thermal equilibrium of the populations is not justified, the attempts to model the pickup ion velocity distribution using the superposition of Maxwell distributions are made by some authors \citep[see, e.g.,][]{zank2010,zirnstein2017,kornbleuth2018}. 

Additionally, we note that the partition of PUIs into two Maxwellian populations is performed only immediately after (downstream) the TS, and in the IHS, the solution (\ref{eq:solution_refl}) of the kinetic equation is used. This method compares favorably with other existing approaches since it does not imply the assumption of fixed temperature fraction in the whole inner heliosheath \citep[like it was done in][]{zirnstein2017, kornbleuth2018}, which is not justified.


\subsection{Comparison of the scenarios}

\begin{figure*}
\includegraphics[width=\textwidth]{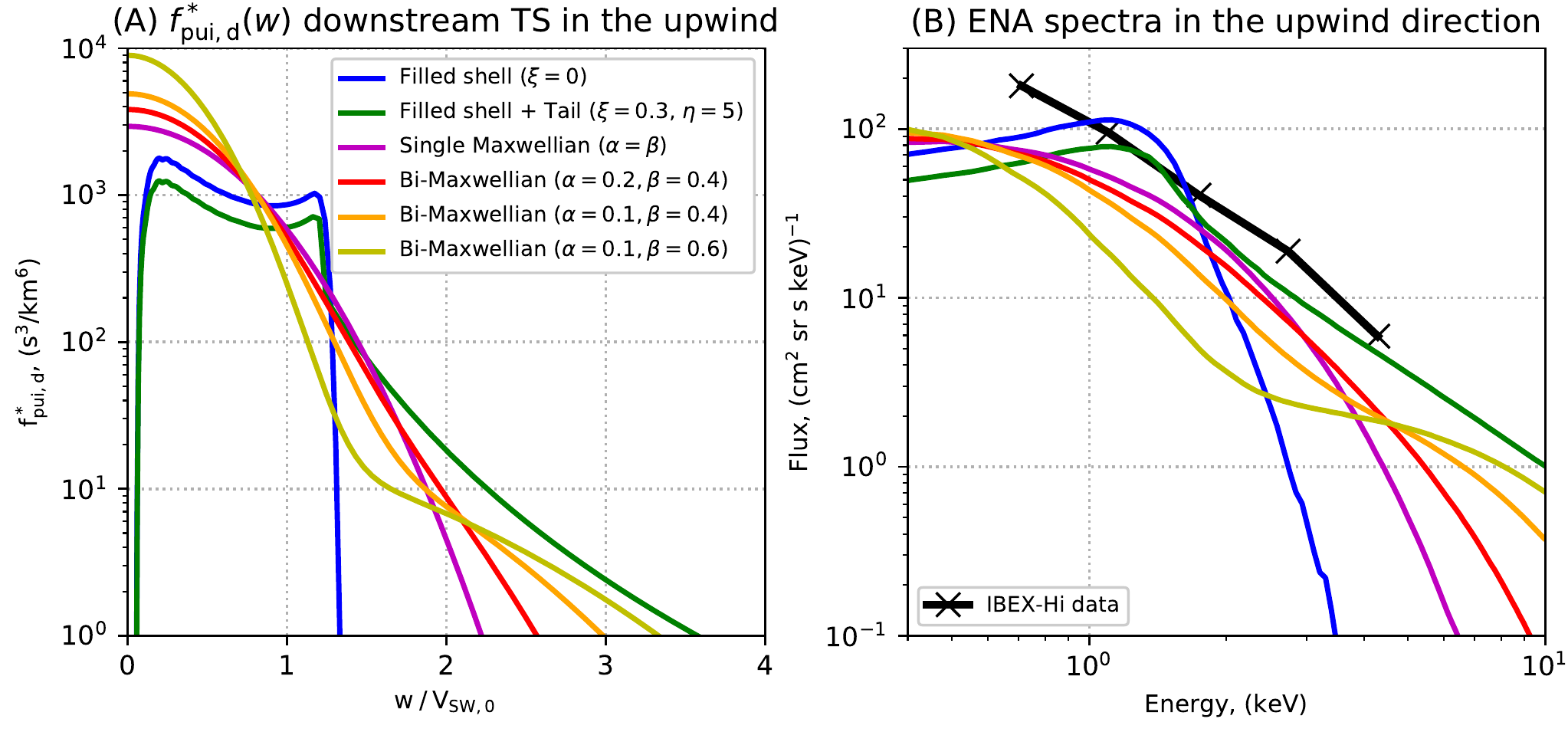}
\caption{
(A) the profiles of the velocity distribution function downstream of the TS and (B) the ENA flux spectra in the upwind direction, simulated using various assumptions on PUI distribution shape immediately downstream of the TS: the blue curves -- non-modified filled shell distribution; the green curves -- <<power-law tail>> scenario (for $\xi = 0.3$ and $\eta = 5$); the magenta curve -- single Maxwellian; the red, orange, and yellow curves -- bi-Maxwellian for different values of $\alpha$ and $\beta$ (shown in the legend). The black solid line with crosses provides the ENA spectra based on the {\it IBEX-Hi} data \citep[taken from][]{schwadron2014}. $V_{\rm sw,0}$ = 432 km s$^{-1}$. 
To convert  the ratio $w / V_{\rm sw,0}$ to energy in the plasma reference frame, the formula $E ({\rm keV}) = 0.974 \times (w / V_{\rm sw,0})^2$ should be used. 
\label{fig:fpui_spectra_upwind}
}
\end{figure*}

Figure \ref{fig:fpui_spectra_upwind} provides the profiles of the velocity distribution function downstream of the TS (plot A) and the ENA flux spectra in the upwind direction (plot B), simulated using various assumptions on PUI velocity distribution. The calculations of the velocity distribution function in Plot A were performed using the plasma and neutral distribution provided by \citet{izmod2020} model in the stationary case. The ENA flux spectra in Plot B were calculated in the time-dependent case and averaged over 2009--2013 (which allows us to compare it to the {\it IBEX-Hi} data directly). 
To plot the {\it IBEX-Hi} ENA spectrum (black solid line with crosses), we use the data extracted from the particular $6^\circ \times 6^\circ$ bin of the full-sky GDF maps presented by \citet{schwadron2014} (more information on the source of the data can be found in Section 3).  The center direction of the chosen bin is (255$^\circ$ in longitude and 3$^\circ$ in latitude in ecliptic J2000 coordinates), which is closest to the upwind direction \citep[255.4$^\circ$, 5.2$^\circ$;][]{witte2004}.
The results of calculations using non-modified filled shell distribution downstream of the TS are presented using the blue curves, the green curves provide the profiles based on the <<power-law tail>> scenario (for $\xi = 0.3$ and $\eta = 5$, see Equations \ref {eq:bound_cond_tail} and \ref{eq:ftail}), the magenta curves -- single Maxwellian distribution (corresponds to the case $\alpha = \beta$), the red, orange, and yellow curves present the results of calculations with reflected particles taken into account, i.e. the bi-Maxwellian distribution (\ref{eq:bound_cond_mxwl}) was utilized with different values of parameters $\alpha$ and $\beta$ (shown in the legend).

As can be seen from Figure \ref{fig:fpui_spectra_upwind}, the power-law <<tail>> in the PUI distribution forms the proper slope of the spectrum and makes the fluxes higher at energies $\gtrsim$ 2 keV. 
The <<maxwellization>> of the PUI distribution function flattens the observed ENA spectrum as well. The red, orange and yellow curves correspond to the cases of $T_{\rm pui}^{\rm ref} / T_{\rm pui} = \beta / \alpha$ ratio equals 2, 4, and 6, respectively. The higher temperature of reflected PUIs the broader <<wings>> in PUI distribution function (ENA spectrum) at high velocities (energies), which also generally leads to a colder temperature of the transmitted particles and, therefore, higher fluxes at small energies ($<$0.7 keV) and smaller fluxes in the {\it IBEX-Hi} energy range ($\sim$0.7--6 keV) in general. Let us also note that the higher the $\beta/\alpha$ ratio, the higher the scaling factor that the model fluxes have to be multiplied by to fit the {\it IBEX-Hi} data.

\section{Fitting the {\it IBEX-Hi} data}\label{sec:fitting}



In our study, we want to characterize the energetic pickup ion population downstream of the TS. For these purposes, we fit the {\it IBEX-Hi} data using our kinetic model described in the previous section. We make use of the GDF data sets provided by \citet{schwadron2014} and obtained by the {\it IBEX-Hi} instrument from 2009 to 2013. These data are available on the webpage of the {\it IBEX} public Data Release 8 (\url{http://ibex.swri.edu/ibexpublicdata/Data_Release_8/}). We model the full-sky ENA flux maps as seen by {\it IBEX-Hi} at 1 au 
in the ram-directions, and with the Compton-Getting and survival probability corrections applied \citep[as it was done in][]{baliukin2020}.


Hereafter we use the notation of vector $\mathbf{p}$ for the pair of parameters, either ($\xi$, $\eta$) or ($\alpha$, $\beta$) depending on the scenario, that defines the energetic pickup proton population downstream of the TS. Let us note that, in principle, the parameters $\mathbf{p}$ rely on the local TS properties, since the efficiency of acceleration processes depends on the compression factor and the shock-normal angle, in particular.
For the specified pair of parameters $\mathbf{p}$, the velocity distribution function of pickup protons in the heliosphere can be calculated.


We perform the parametric study by varying parameters $\mathbf{p}$ in the wide range of values to minimize the difference between the model results and data, which can be described in terms of the reduced chi-square statistic that is $\chi^2$ per degree of freedom:
\begin{equation}
	\chi^2_{\rm red}(\mathbf{p}, k) = \frac{1}{N - M} \sum_{\rm ESA_i} \sum_{\rm LOS_j} \left( \frac{J^{\rm d}_{\rm i j} - k \cdot J^{\rm m}_{\rm i j}(\mathbf{p})}{\sigma_{\rm i j}} \right)^2,
\label{eq:chi2}
\end{equation}
where $k$ is the scaling coefficient, $J^{\rm d}_{\rm i j}$ and $J^{\rm m}_{\rm i j}$ are the data and model values of the ENA flux, $\sigma_{\rm i j}$ are the uncertainties of the data. The index <<i>> identifies the {\it IBEX-Hi} energy channel ($i = 2,\ldots ,6$, i.e. the top five of them are considered), and the index <<j>> represents the line-of-sight $\rm LOS_j$ that belongs to the considered region of the sky that will be described later (the 6-degree binned {\it IBEX-Hi} data of the full-sky contains $60 \times 30 = 1800$ lines-of-sight). $N$ is the number of observations in 5 energy channels and within the considered region, $M$ is the number of fitted parameters, so the number of degrees of freedom is $N - M$. Let us additionally note that model values of the fluxes $J^{\rm m}_{\rm i j}$ were calculated using the time-dependent model and averaged over 2009-2013 (in correspondence with the data we compare with).

The weighted linear regression provides the best-fitting value $\hat k$ that minimizes the $\chi^2_{\rm red}(k)$ statistic
\begin{equation}
	\hat k = \frac{\sum_{\rm i j} J^{\rm m}_{\rm i j} J^{\rm d}_{\rm i j} / \sigma_{\rm i j}^2 }{\sum_{\rm i j} (J^{\rm m}_{\rm i j} / \sigma_{\rm i j})^2},
\label{eq:scaling_factor}
\end{equation}
Hereafter, the hat over the variable represents its best-fitting value.

\begin{figure*}
\includegraphics[width=1.1\textwidth]{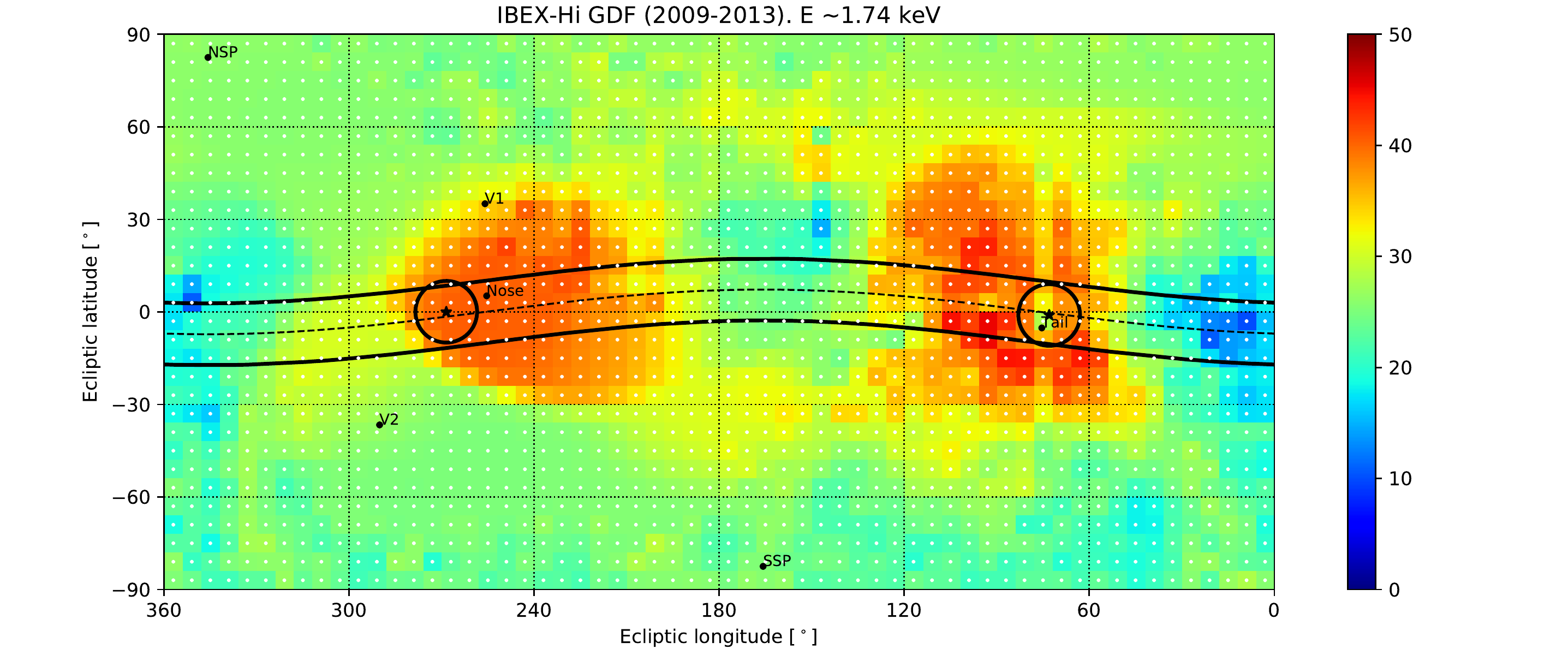}
\caption{The fitting regions of the sky used on different steps. White dots represent centers of the bins for which the {\it IBEX-Hi} data is presented. The figure shows the GDF data from energy channel 4 collected during 2009-2013. Black circles show upwind and downwind 10$^\circ$ cones (black stars represent its centers), and the lines-of-sight that belong to these cones are used in Step 1 of the fitting procedure. Black dashed line is the solar equatorial plane. The swath of the sky between the black solid lines, which consist from the lines-of-sight that are less than 10$^\circ$ away from the solar equatorial plane, is used for fitting on Step 2. The units of fluxes are $(\rm cm^2\: sr\: s\: keV)^{-1}$.
}
\label{fig:fit_regions}
\end{figure*}

We split our fitting procedure into four consecutive steps:
\begin{enumerate}

\item Fitting the parameters in the Nose and Tail regions. In this step, we select those regions of the sky, where the streamlines are almost straight. The ENA spectra for these directions reflect properties of the energetic pickup population only from the same region at the TS, so it is possible to probe the parameters of energetic PUIs directly. The directions from the Nose and Tail regions, along which the streamlines are almost straight, were obtained in the frame of the stationary heliospheric model, and they are (268.5$^\circ$, 0.0$^\circ$) and ($73.0^\circ$, $-1.0^\circ$) in ecliptic (J2000) coordinates, which are shifted by $\sim$14$^\circ$ and $\sim$5$^\circ$ from the upwind and downwind directions, respectively. After that, we select the lines-of-sight of the {\it IBEX-Hi} binned data that are less than 10$^\circ$ away from the straight streamlines in the Nose and Tail regions, i.e., belong to the 10$^\circ$ cones (see Figure \ref{fig:fit_regions}). The results of the fitting are weakly dependent on the size of the cone (5$^\circ$ and 20$^\circ$ were also tested).

Since the fluxes in the Nose and Tail cones depend only on the parameters of energetic PUIs in the corresponding regions of the TS and the cones do not intersect, we assume that the model fluxes in the Nose cone depend only on two constant parameters $\mathbf{p}_{\rm upw}$ that reflect properties of energetic pickup ions in the upwind region of the TS, and the fluxes in the Tail cone depend on the pair of constant parameters $\mathbf{p}_{\rm dwd}$, which defines the energetic population in the downwind region. Therefore, the reduced chi-square statistic can be decomposed into two sums:
\begin{equation}
\begin{split}
&\chi^2_{\rm red}(\mathbf{p}_{\rm upw},\: \mathbf{p}_{\rm dwd}, k) = \frac{1}{N - M} \times \\
&\sum_{\rm ESA_i} \left( \sum_{\rm Nose} \left( \frac{J^{\rm d}_{\rm i j} - k \cdot J^{\rm m}_{\rm i j}(\mathbf{p}_{\rm upw})}{\sigma_{\rm i j}} \right)^2 + \sum_{\rm Tail} \left( \frac{J^{\rm d}_{\rm i j} - k \cdot J^{\rm m}_{\rm i j}(\mathbf{p}_{\rm dwd})}{\sigma_{\rm i j}} \right)^2 \right),
\end{split}
\end{equation}
where the number of observations $N = 80$ (5 channels and 16 lines-of-sight, 8 in the Nose and 8 in the Tail regions), the number of model parameters $M = 5$ (2 parameters for each of the cones, and common scaling coefficient $k$), and the summations are performed for the lines-of-sight $\rm LOS_j$ within the Nose and Tail cones. The best-fitting scaling coefficient $\hat k = \hat k(\mathbf{p}_{\rm upw},\:\mathbf{p}_{\rm dwd})$ can be calculated using Equation (\ref{eq:scaling_factor}).

We have varied 4 parameters, $\mathbf{p}_{\rm upw}$ and $\mathbf{p}_{\rm dwd}$, in the wide range, and, as the result of this step, its best-fitting values, as well as the scaling coefficient $\hat k$ that minimizes the chi-square statistic, were obtained. These parameters are used in the next steps.

\item Fitting the parameters at the flanks of the heliosphere. The effectiveness of proton acceleration at the TS depends on its local properties and, in particular, on the shock-normal angle \citep[between the magnetic field vector and normal to the shock surface; see, e.g.,][]{chalov2015}. At the flanks of the heliosphere, the shock-normal angle is significantly different (as small as $\approx$ 70$^\circ$) compared to the Nose, Tail, and pole regions, where this angle is $\approx$ 90$^\circ$ \citep[see, e.g., Figure 2B in][]{baliukin2020}. Let us note, that the dependence of the velocity distribution function of pickup protons downstream of the TS on the shock-normal angle was also taken into account through the jump condition at the TS \citep[see Equation 14 in][]{baliukin2020}.

To introduce the influence of the shock-normal angle variation along the TS on the energetic PUI properties we set $\mathbf{p}_{\rm flank}$ at the flanks of the TS as free parameters and specify the dependence of $\mathbf{p}$ on heliolongitude $\varphi_{\rm helio} \in [0, 2 \pi)$, which is counted from X-axis in XY (solar equatorial) plane of the heliographic inertial (HGI) coordinate system. We assign parameters $\mathbf{p}_{\rm upw}$ estimated at the Nose region to the direction ($\lambda_{\rm helio} = 0$, $\varphi_{\rm helio} = \pi$) that is close to the upwind direction, and parameters $\mathbf{p}_{\rm dwd}$ -- to the direction ($\lambda_{\rm helio} = 0$, $\varphi_{\rm helio} = 0$), where $\lambda_{\rm helio} \in [-\pi/2,\: \pi/2]$ is the heliolatitude counted from the solar equatorial plane ($\lambda_{\rm helio} = 0$). The parameters $\mathbf{p}_{\rm flank}$ are responsible for the directions ($\lambda_{\rm helio} = 0$, $\varphi_{\rm helio} = \pi/2$) and ($\lambda_{\rm helio} = 0$, $\varphi_{\rm helio} = 3\pi/2$).

In the solar equatorial plane we consider a piecewise linear dependence on heliolongitude. To be more precise,  
\begin{equation}
	\mathbf{p}_{\rm seq} =
	\begin{cases}
		\mathbf{p}_{\rm upw} + 2 \zeta (\mathbf{p}_{\rm flank} - \mathbf{p}_{\rm upw}), 			& 0 \leq \zeta < 1/2, \\
		\mathbf{p}_{\rm dwd} + 2 (1 - \zeta) (\mathbf{p}_{\rm flank} - \mathbf{p}_{\rm dwd}), 	& 1/2 \leq \zeta \leq 1,
	\end{cases}
\label{eq:heliolon}
\end{equation}
where $\zeta = \lvert 1 - \varphi_{\rm helio} / \pi \rvert$. At this step, we assume that parameters at the TS depend only on heliolongitude (and not on heliolatitude), i.e., $\mathbf{p} = \mathbf{p}_{\rm seq}(\mathbf{p}_{\rm upw}, \mathbf{p}_{\rm dwd}, \mathbf{p}_{\rm flank}, \varphi_{\rm helio})$. Therefore, the number of model parameters $M = 7$, which are $\mathbf{p}_{\rm upw}$, $\mathbf{p}_{\rm dwd}$, $\mathbf{p}_{\rm flank}$, and $k$.

We assume that the ENA fluxes in the directions of a narrow swath in the solar equatorial plane proximity depend on the TS properties only from this swath of the TS.
For the fitting, the lines-of-sight that are inclined by less than 10$^\circ$ to the solar equatorial plane were selected (206 lines-of-sight satisfy this condition, 5 energy channels, and, therefore, $N = 1030$). By varying the parameters $\mathbf{p}_{\rm flank}$ and minimizing the chi-square statistic, its best-fitting values were estimated. These values are used in the next steps.

\item Fitting the parameters at the solar poles. In this step, we introduce the dependence of energetic PUI parameters at the TS with the heliolatitude $\lambda_{\rm helio}$, which can be induced by the heliolatitude dependence of the solar wind (slow SW at the equatorial plane versus fast SW at the poles). We set $\mathbf{p}_{\rm pole}$ as free parameters at the poles ($\lambda_{\rm helio} = \pm \pi/2$), so the general form of dependence can be written as $\mathbf{p} = \mathbf{p}(\mathbf{p}_{\rm seq}(\varphi_{\rm helio}), \mathbf{p}_{\rm pole}, \lambda_{\rm helio})$. Since there is no additional information on the dependence of these parameters on heliolatitude, for the sake of simplicity we assume it to be linear:
\begin{equation}
	\mathbf{p} = \mathbf{p}_{\rm seq} + \frac{2 \lvert\lambda_{\rm helio}\rvert}{\pi} (\mathbf{p}_{\rm pole} - \mathbf{p}_{\rm seq}).
\label{eq:heliolat}
\end{equation}

At this step, the parameters of the energetic population depend both on heliolongitude and heliolatitude linearly, and the number of model parameters $M = 9$: $\mathbf{p}_{\rm upw}$, $\mathbf{p}_{\rm dwd}$, $\mathbf{p}_{\rm flank}$, $\mathbf{p}_{\rm pole}$, and $k$. We  fit the full-sky {\it IBEX-Hi} data, so the number of observations $N = 9000$ (5 energy channels and 1800 lines-of-sight). As the result of this step, the parameters $\mathbf{\hat p}_{\rm pole}$ were estimated.

\item Variation of the parameters ($\mathbf{p}_{\rm upw}$, $\mathbf{p}_{\rm dwd}$, $\mathbf{p}_{\rm flank}$, $\mathbf{p}_{\rm pole}$) iteratively until the convergence is observed. To be more precise, we have varied each pair of parameters one by one while the other parameters were assumed constant. Therefore, at this step, the number of model parameters is still $M = 9$. The full-sky ENA maps were fitted, so the number of observations is $N = 9000$. In our simulations, 5 iterations were needed to obtain the convergence.

\end{enumerate}

The fitting algorithm described above allows obtaining the approximate solution after the first three steps (on each of them the model fluxes qualitatively depend only on two parameters) and, therefore, significantly reduces the number of simulations needed to obtain the best-fitting parameters.

Let us additionally note that the algorithm assumes the linear dependence of parameters, which define the energetic population of pickup protons downstream of the termination shock, both on heliolongitude and heliolatitude (see Equations \ref{eq:heliolon} and \ref{eq:heliolat}), while its time dependence is not considered. The time dependence can be caused by the variable solar wind conditions that influence the parameters, especially at the poles, where the SW velocity changes with time significantly. In this work, we make use of the data that was averaged over 2009 -- 2013, so our results represent some kind of mean over this time period.

\section{Results} \label{sec:results}

The procedure described in the previous section was applied to both scenarios of the velocity distribution function downstream of the TS.

\subsection{<<Power-law tail>> scenario}

\begin{table*}
	\centering
	\caption{Best-fitting parameters for <<power-law tail>> scenario of modification the PUI distribution function downstream of the termination shock.}
	\begin{tabular}{llcccccccccccc}
		\hline
		\hline
		Step & Fitting region 	& $\hat \xi_{\rm upw}$ & $\hat \xi_{\rm dwd}$ & $\hat \xi_{\rm flank}$ & $\hat \xi_{\rm pole}$ & $\hat \eta_{\rm upw}$ & $\hat \eta_{\rm dwd}$ & $\hat \eta_{\rm flank}$  & $\hat \eta_{\rm pole}$ & $\hat k$ & $N$ & $M$ & $\chi^2_{\rm red, min}$\\
		\hline
		1 & Nose and Tail 		& 0.27 & 0.60 & - 	& - 		& 5.0 & 3.1 & - 		& - 		& 1.55 	& 80 	& 5	& 8.72 \\
		2 & Solar equator swath	& 0.27 & 0.60 & 0.45 	& - 		& 5.0 & 3.1 & 3.2 	& - 		& 1.62 	& 1030 	& 7 	& 12.66 \\
		3 & Full-sky 			& 0.27 & 0.60 & 0.45 	& 0.68 	& 5.0 & 3.1 & 3.2 	& 3.4 	& 1.57 	& 9000 	& 9 	& 18.03 \\
		\hline
		4 & Full-sky 			& 0.22 & 0.68 & 0.42 	& 0.67 	& 5.3 & 3.01 & 3.3 	& 3.4 	& 1.54 	& 9000 	& 9 	& 17.05 \\
	\end{tabular}
	\label{tab:fitting_powerlaw}
\end{table*}

Table \ref{tab:fitting_powerlaw} summarizes the best-fitting parameters obtained on each of the four steps using the model with the <<power-law tail>> scenario for the energetic pickup proton population downstream of the TS (see Section \ref{sec:powertail}). The figures that justify these estimations can be found in Section \ref{app:powertail} of Appendix \ref{app:fitting_steps}. To obtain the best consistency with the data, the modelled ENA fluxes should be multiplied by a factor of $\hat k (\hat \xi,\: \hat \eta) = 1.54$.

\begin{figure}
\includegraphics[width=\columnwidth]{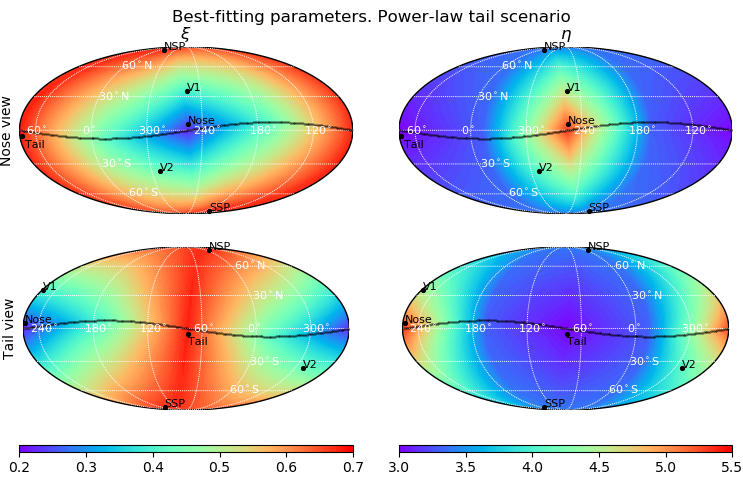}
\caption{Distribution of best-fitting parameters $\xi$ and $\eta$ (<<power-law tail>> scenario) downstream of the  termination shock depending on the line-of-sight. The maps are in ecliptic coordinates (J2000). The black line shows the solar equator. The maps in the top and bottom rows are the same but are centered in different longitudes (upwind and downwind, respectively). NSP = north solar pole, SSP = south solar pole.
}
\label{fig:powtail_parameters}
\end{figure}

Figure \ref{fig:powtail_parameters} shows the distribution of the best-fitting parameters $\xi$ and $\eta$ downstream of the termination shock in the form of the full-sky maps in the ecliptic coordinates. As can be seen, the density fraction of the tail distribution $\xi$ is the lowest ($\approx$ 20\%) in the upwind and the highest ($\approx$ 70\%) in the downwind and pole directions. The spectral index $\eta$ has its maximum in the upwind, so the energy spectrum of pickup protons downstream of the TS is relatively soft in the Nose region. The best-fitting value of the spectral index in the upwind direction is $\hat \eta_{\rm upw} = 5.3$, which is rather close to the value 5 reported by \citet{fisk2007}. A hard energy spectrum of PUIs (with the spectral index $\eta \approx$ 3 -- 3.5) is seen in the whole downwind hemisphere of the TS. It should be noted that the <<power-law tail>> scenario described in Section \ref{sec:powertail} implies that $\eta > 3$. Nevertheless, the fitting procedure suggests that the spectral index in the downwind direction tends to be 3 ($\eta_{\rm dwd} \to 3$).

\subsection{<<Bi-Maxwellian>> scenario}

\begin{table*}
	\centering
	\caption{Best-fitting parameters for <<bi-Maxwellian>> scenario of modification the PUI distribution function downstream of the termination shock.}
	\begin{tabular}{llcccccccccccc}
		\hline
		\hline
		Step & Fitting region 	& $\hat \alpha_{\rm upw}$ & $\hat \alpha_{\rm dwd}$ & $\hat \alpha_{\rm flank}$ & $\hat \alpha_{\rm pole}$ & $\hat \beta_{\rm upw}$ & $\hat \beta_{\rm dwd}$ & $\hat \beta_{\rm flank}$  & $\hat \beta_{\rm pole}$ & $\hat k$ & $N$ & $M$ & $\chi^2_{\rm red, min}$\\
		\hline
		1 & Nose and Tail 		& 0.26 & 0.06 & - 	& - 		& 0.48 & 0.52 & - 	& - 		& 2.30 	& 80 	& 5	& 5.53 \\
		2 & Solar equator swath 	& 0.26 & 0.06 & 0.05 	& - 		& 0.48 & 0.52 & 0.30 	& - 		& 2.37 	& 1030	& 7 	& 10.62 \\
		3 & Full-sky 			& 0.26 & 0.06 & 0.05 	& 0.20 	& 0.48 & 0.52 & 0.30 	& 0.63 	& 2.44 	& 9000 	& 9	& 12.28 \\
		\hline
		4 & Full-sky 		& 0.41 & 0.04 & 0.03 & 0.19 	& 0.62 & 0.70 & 0.21 & 0.61 	& 2.43 	& 9000 	& 9	& 7.77 \\
	\end{tabular}
	\label{tab:fitting_bimaxwl}
\end{table*}

The best-fitting parameters obtained in the frame of the <<bi-Maxwellian>> scenario for the energetic PUI population downstream of the TS (see Section \ref{sec:bimaxwellian}) for all four steps of the fitting procedure are compiled in Table \ref{tab:fitting_bimaxwl}. The figures that justify the fitting steps are shown in Section \ref{app:bi-maxwl} of Appendix \ref{app:fitting_steps}. The best agreement with the data can be achieved if our modeling results are scaled by a factor of $\hat k (\hat \alpha,\: \hat \beta) = 2.43$, which is substantially higher compared to the <<power-law tail>> scenario (1.53).

\begin{figure}
\includegraphics[width=\columnwidth]{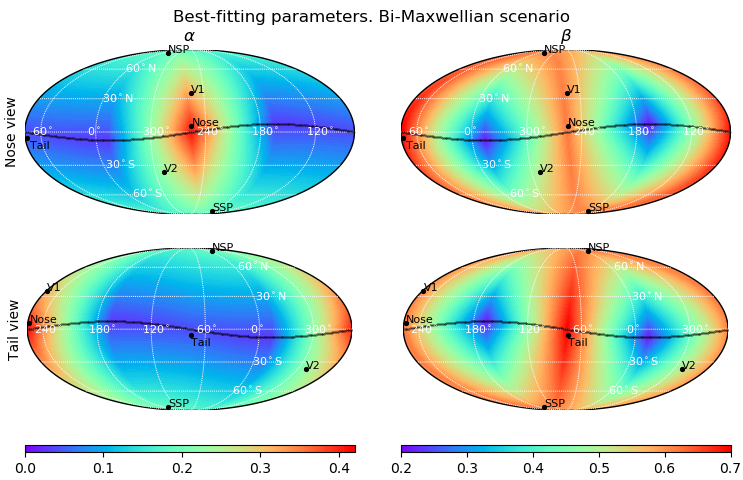}
\caption{
The description is the same as for Figure \ref{fig:powtail_parameters}, but the best-fitting parameters $\alpha$ and $\beta$ of the <<bi-Maxwellian>> scenario are presented.
}
\label{fig:bimaxwl_parameters}
\end{figure}

The values of the parameters $(\alpha,\: \beta)$, which were usually assumed in the previous works on modelling, are $\alpha = 0.08$ \citep{chalov1996,zank2010} and $\beta = 0.479$ \citep{zirnstein2017}. In Figure \ref{fig:bimaxwl_parameters} the maps (in ecliptic coordinates) of the best-fitting parameters $\alpha$ and $\beta$ downstream of the TS are shown. The density fraction of the reflected PUI population is maximal ($\approx$ 40\%) in the Nose region, while in the downwind direction and at the flanks of the TS this population accounts for only a few percent of the total number of PUIs. The highest thermal pressure fraction $\beta$ is seen in the downwind direction, being rather high at the poles and in the upwind direction. The $\beta/\alpha$ ratio is maximal in the downwind direction also, so the reflected PUIs are heated in this region the most.


As seen from Tables \ref{tab:fitting_powerlaw} and \ref{tab:fitting_bimaxwl}, the $\chi^2_{\rm red}$ statistic is increasing during the first three steps, which can be confusing at the first glance. It can be explained by different fitting regions (with the increasing number of observations $N$) considered on each of these steps. The values of the $\chi^2_{\rm red,min}$ obtained at the final fourth step in both scenarios are sufficiently larger than 1. That is generally considered as an indicator of underestimation of the data uncertainties or model underfitting, which can be induced by a more complex distribution of energetic PUIs downstream the TS than it is assumed in the model. The other source of the model nonconformity might be the assumptions made in the fitting algorithm, such as the heliolatitude and heliolongitude linear dependence and time independence of the parameters.

\subsection{Velocity distribution of pickup protons for different locations downstream of the termination shock} \label{sec:vdf}

\begin{figure*}
\includegraphics[width=\textwidth]{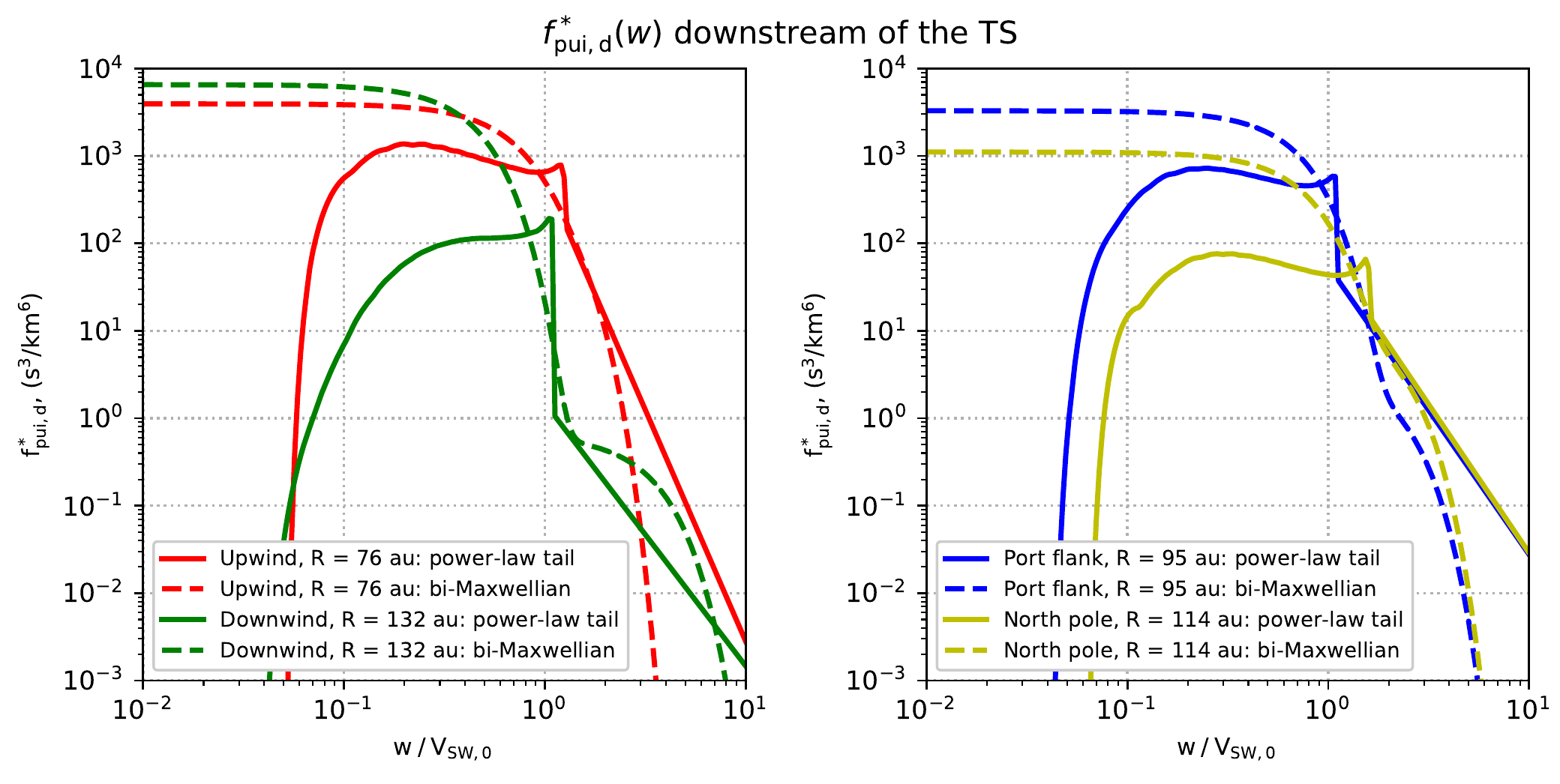}
\caption{The velocity distribution function of PUIs downstream of the TS, which was calculated under two scenarios and for different directions: solid lines -- <<power-law tail>>, dashed lines -- <<bi-Maxwellian>>. The modelling was performed using the stationary heliospheric model and best-fitting parameters of the energetic PUI population, which are compiled in Tables \ref{tab:fitting_powerlaw} and \ref{tab:fitting_bimaxwl}. No scaling was applied. $V_{\rm sw,0}$ = 432 km s$^{-1}$.
To convert  the ratio $w / V_{\rm sw,0}$ to energy in the plasma reference frame, the formula $E ({\rm keV}) = 0.974 \times (w / V_{\rm sw,0})^2$ should be used.
}
\label{fig:fpui_TS}
\end{figure*}

Figure \ref{fig:fpui_TS} shows the velocity distribution function of PUIs downstream of the TS that was calculated under two scenarios (solid lines -- <<power-law tail>>, dashed lines -- <<bi-Maxwellian>>) and in four selected directions (upwind, downwind, port flank, and north solar pole). The distances to the TS in these directions are 76, 132, 95, and 114 au, respectively, which were calculated in the frame of the stationary version of the \citet{izmod2020} model.
The port flank is on the left side of the heliosphere (with respect to the upwind direction) as seen from the Sun. The modelling of the velocity profiles in Figure \ref{fig:fpui_TS} was performed using the best-fitting parameters of the energetic population, which are compiled in Tables \ref{tab:fitting_powerlaw} and \ref{tab:fitting_bimaxwl}, and based on the stationary version of the heliospheric model. Let us note that the velocity distribution functions are presented without any scaling applied, while the results of the fitting procedure indicate that the ENA fluxes should be scaled (by the factors 1.54 and 2.43 in the frame of the <<power-law tail>> and <<bi-Maxwellian>> scenarios, respectively).

As can be seen from Figure \ref{fig:fpui_TS}, two scenarios produce consistent results in the range of velocities $1 \lesssim w / V_{\rm SW,0} \lesssim 3$. The PUIs with such velocities cover the {\it IBEX-Hi} energy range of the ENA flux spectrum, as seen from the comparison of Plots A and B in Figure \ref{fig:fpui_spectra_upwind}. The <<bi-Maxwellian>> scenario produces more pickup protons with velocities $w / V_{\rm SW,0} \lesssim 1$, while the <<power-law tail>> scenario generally provides more PUIs with high velocities ($w / V_{\rm SW,0} \gtrsim 3$).

The PUI energy spectrum is hardest in the direction of the solar poles (yellow curves in Figure \ref{fig:fpui_TS}), which is also observed in the {\it IBEX-Hi} data of ENA fluxes \cite[see, e.g., Figure 8 in][]{schwadron2014}. Our results support the conclusion of hybrid numerical simulations performed by \citet{giacalone2021} that pickup protons are heated more across the TS in the tail than at other locations, as can be seen from the comparison of the green curves with other lines. For the downwind direction, the spectral index $\eta$ is the lowest and the $\beta/\alpha$ ratio is the highest. Therefore, the acceleration is the most effective in the tail region.

\subsection{Full-sky ENA flux maps} \label{sec:fullsky_maps}

\begin{figure*}
\includegraphics[width=\textwidth]{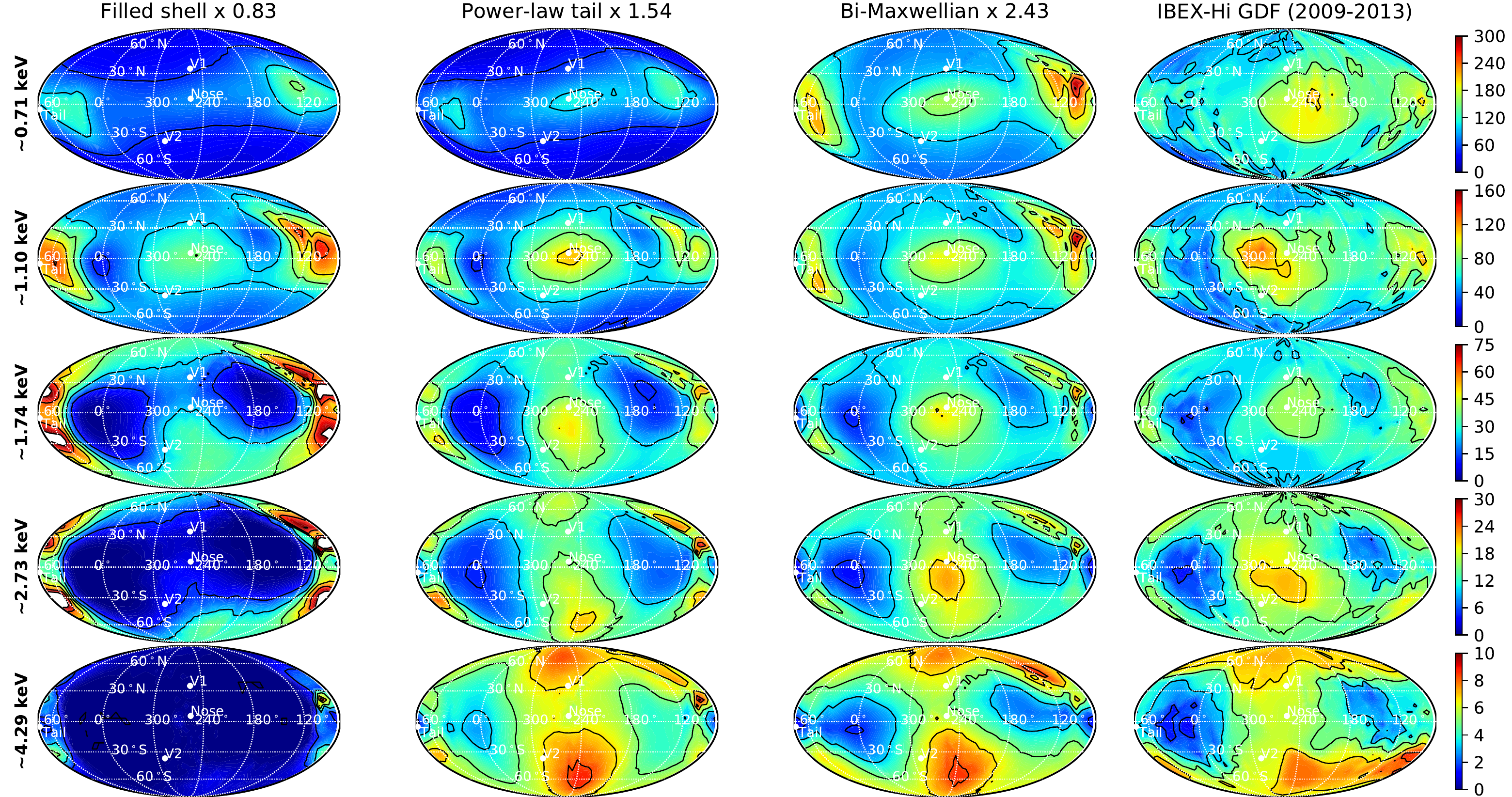}
\caption{
Full-sky ENA flux maps (the Mollweide projections) in ecliptic (J2000) coordinates as it was observed by {\it IBEX-Hi} at the energy channels 2--6 (by rows). The first column shows the results of calculations without additional energetic population taken into account (filled shell distribution downstream of the TS). The second and third columns present the results of calculations using <<power-law tail>> and <<bi-Maxwellian>> scenarios for the distribution function downstream of the TS (with best-fitting parameters applied), respectively. The fourth column presents the {\it IBEX-Hi} data collected during 2009 -- 2013. The units of fluxes are $(\rm cm^2\: sr\: s\: keV)^{-1}$. The white color is due to exceeding the upper limit of flux intensity. The maps are centered on the upwind longitude $255.4^\circ$.}
\label{fig:maps_nose}
\end{figure*}

\begin{figure*}
\includegraphics[width=\textwidth]{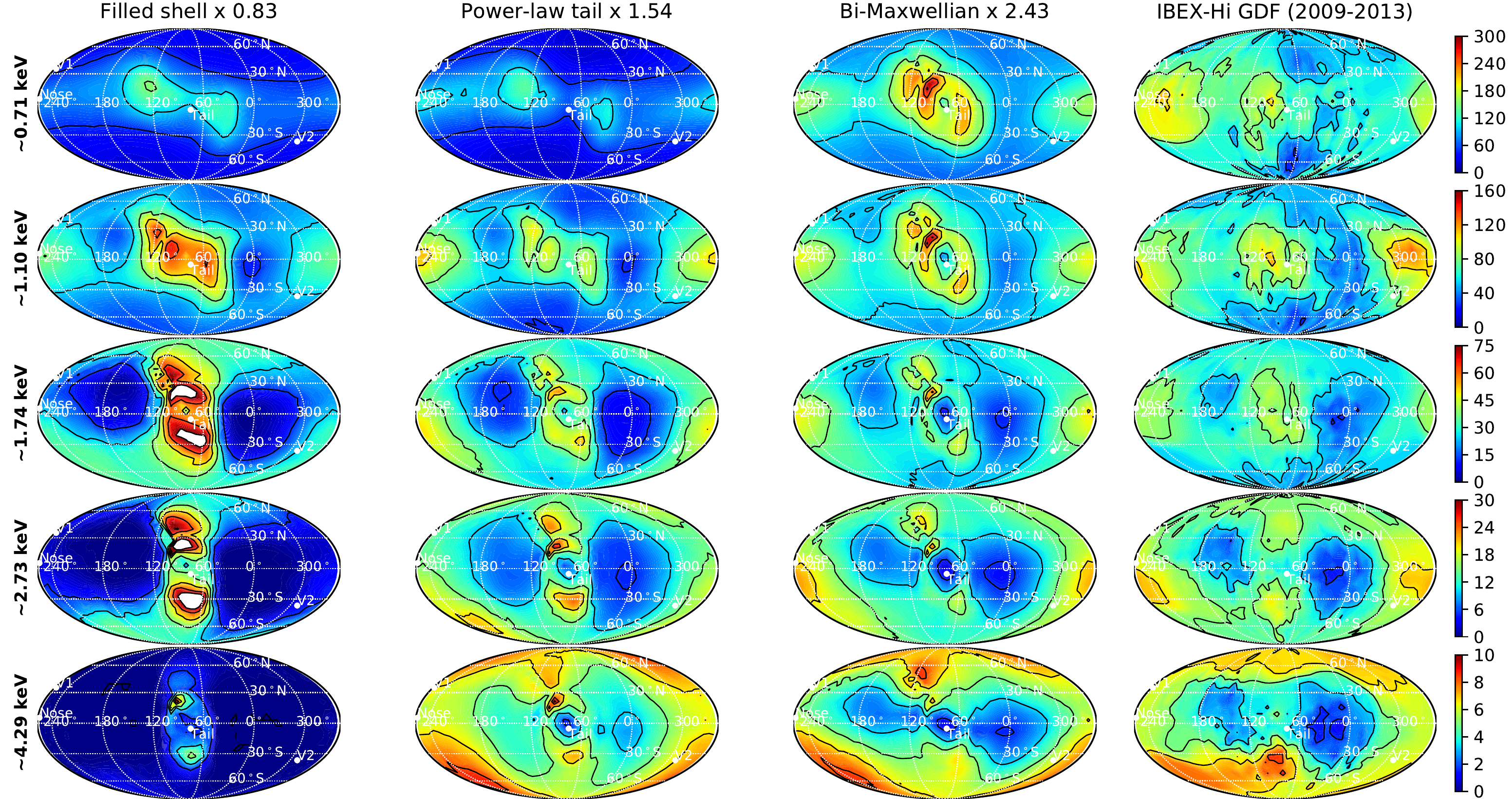}
\caption{The description is the same as for Figure \ref{fig:maps_nose}, but the maps are centered on the downwind longitude $75.4^\circ$.}
\label{fig:maps_tail}
\end{figure*}

Figures \ref{fig:maps_nose} and \ref{fig:maps_tail} show the {\it IBEX-Hi} data of the top five energy channels and the simulated full-sky maps of the ENA fluxes. The maps in Figures \ref{fig:maps_nose} and \ref{fig:maps_tail} are the same and cover the whole full-sky but are centered in different longitudes (upwind and downwind, respectively). The first columns of the figures present the results of calculations without the energetic population taken into account, i.e., the filled shell distribution downstream of the TS was assumed ($\xi = 0$). The appropriate scaling coefficient, in this case, is 0.83, and it was applied to the fluxes shown in the first columns. The second and third columns show the results of calculations using the <<power-law tail>> and <<bi-Maxwellian>> scenarios with the best-fitting parameters applied. The fourth columns present the {\it IBEX-Hi} data obtained in 2009 -- 2013.

As seen from Figures \ref{fig:maps_nose} and \ref{fig:maps_tail}, the modelling results with energetic PUI population taken into account (second and third columns) reproduce the data both qualitatively and quantitatively well. All the main features of the flux maps, such as the maximum in the Nose region, the low flux lobes at the flanks of the heliosphere, and the North/South heliotail lobes, which are present in the {\it IBEX-Hi} data, are observed in the model calculations as well. The energetic population of PUIs (modelled both in the frame of the <<power-law tail>> and <<bi-Maxwellian>> scenarios) makes the model results considerably more consistent with the data, since it provides higher fluxes at $\sim$ 2.73 and $\sim$4.29 keV energy channels. Therefore, we can conclude that the globally distributed flux is highly sensitive to the shape of the velocity distribution function of PUIs in the inner heliosheath and accounting for the existence of the energetic population of pickup protons is vital to explain the data. 

The <<bi-Maxwellian>> scenario provides higher fluxes from the Nose and Tail regions of the sky map at the lowest energy channel ($\sim$0.71 keV), which is better consistent with the data compared to the case of the <<power-law tail>> scenario. As can be seen in Figure \ref{fig:fpui_spectra_upwind}, the <<maxwellization>> flattens the energy spectrum and redistribute fluxes to the lower energy channel. Even though the bi-Maxwellian distribution of PUIs is not physically justified, this scenario reproduces the actual shape of the velocity distribution of pickup protons downstream of the TS better, since the minimum value of the chi-squared reduced statistic is smaller (7.77 versus 17.05 for the <<power-law tail>> scenario, see Tables \ref{tab:fitting_powerlaw} and \ref{tab:fitting_bimaxwl}). At the energy channels $\sim$1.74 keV and higher, both scenarios produce comparable ENA flux maps. 

It is seen in Figure \ref{fig:maps_nose} that in the case of the <<power-law tail>> scenario the maximum of the fluxes in the upwind direction splits up into two peaks at the energy channel $\sim$2.73 keV, while it is undivided for these energies in the results of the <<bi-Maxwellian>> scenario and the {\it IBEX-Hi} data also. This fact indicates that the ENA spectrum in the middle range of {\it IBEX-Hi} energies has a more convex shape than it is predicted by a power-law tail distribution of energetic PUIs (compare solid and dashed red curves in Figure \ref{fig:fpui_TS}).

The shape and depth of the low flux lobes at the flanks of the heliosphere are better predicted by the <<bi-Maxwellian>> scenario. As can be noted, at the flanks the <<power-law tail>> scenario provides lower fluxes at $\sim$1.1 and $\sim$1.74 keV energy channels and higher fluxes at $\sim$4.29 keV, compared to the {\it IBEX-Hi} data. This behavior exhibits the fact that the actual velocity distribution of PUIs is smoother and has a higher spectral index in the range of velocities $1 \lesssim w / V_{\rm SW,0} \lesssim 2$, like it is in the <<bi-Maxwellian>> case (see solid and dashed blue curves in Figure \ref{fig:fpui_TS}).

Important to note that in the <<bi-Maxwellian>> case the fluxes must be scaled (multiplied) by a factor of 2.43 to achieve the obtained level of conformity. By this criterion, the <<power-law tail>> scenario compares favorably (scaling coefficient is 1.54) with the case of superposition of Maxwellian distributions. Nevertheless, for quantitative agreement with the data, the model fluxes should be substantially increased in both scenarios,  which may be an indicator of a lower hydrogen number density used in the heliospheric model than it is in reality.
The H number density in the LISM assumed in \citet{izmod2020} model, which we employ in this work, is $n_{\rm H,LISM}$ = 0.14 cm$^{-3}$. In the upwind direction, the H number density is increasing from $\sim$0.1 cm$^{-3}$ at the TS to $\sim$0.12 cm$^{-3}$ at the HP. It should be noted that the recent analysis of the pickup proton measurements from the SWAP instrument on New Horizons indicates a higher number density of hydrogen atoms than widely used in the models \citep[$\sim$0.127 cm$^{-3}$ at the TS;][]{swaczyna2020}, which is also supported by the analysis of the combination of 5.2--55 keV Cassini/INCA ENAs and >28 keV Voyager/LECP ion measurements \citep{dialynas2019} and by the observations of the interplanetary backscattered Lyman-$\alpha$ emission from different spacecraft \citep[see, e.g.,][]{katushkina2016}. The other possible aspects that may affect the values of the fluxes, as suggested by \citet{zirnstein2017}, are (a) the IHS thickness, which is even larger in the model than suggested by Voyager 1 \& 2 observations, and (b) the velocity diffusion, not considered in our modelling.

\section{Conclusions and discussion} \label{sec:conclusions}

In this work, we have simulated the ENA fluxes as it was observed by the {\it IBEX-Hi} instrument using different assumptions on the PUI velocity distribution downstream of the TS and performed the study of the energetic pickup proton population and its parameters based on the {\it IBEX-Hi} data. The main results of this work can be summarized as follows.

\begin{enumerate}

	\item The kinetic model of pickup ion distribution in the heliosphere developed by \citet{baliukin2020}, which utilizes the results of calculations (plasma and neutral distributions) of the Moscow heliospheric model \citep{izmod2020}, was extended by the introduction of the energetic population of PUIs in the inner heliosheath. For these purposes, two scenarios of the velocity distribution of pickup protons downstream of the heliospheric termination shock were considered: (1) a compressed filled shell distribution with energetic power-law tail ($f_{\rm tail}^{*}(w) \propto w^{-\eta}$), and (2) a bi-Maxwellian distribution. 
	
	\item For both of the considered scenarios, the parametric study was performed with the help of the {\it IBEX-Hi} globally distributed flux data \citep{schwadron2014}, and the best-fitting parameters of the energetic PUI population were determined for different locations at the TS (in the upwind, downwind, heliospheric flank, and solar pole directions). We concluded that the PUI energy spectrum is the hardest in the direction of the solar poles, and the pickup protons are heated more across the TS in the tail than at other locations, so the acceleration is the most effective in this region. The quantitative estimates for the parameters of the energetic population of pickup protons obtained in this work can be useful for testing and verifying other hybrid models that simulate the acceleration of protons at the heliospheric termination shock. 

	\item The ENA flux maps simulated using the model, which takes into account the energetic PUI population, are in qualitative and quantitative agreement with the data. All the main features of the flux maps, which are present in the {\it IBEX-Hi} data, are observed in the model results as well. The population of energetic PUIs increases the ENA fluxes at the top energy channels of the {\it IBEX-Hi} from all across the sky, so the model simulations become consistent with the data.		
	
	\item Some differences between the modelling results and data are not explained by the scenarios considered in our work. For quantitative agreement with the data, the model fluxes should be significantly scaled (multiplied by 1.54 and 2.43 in the <<power-law tail>> and <<bi-Maxwellian>> scenarios, respectively). This may be an indicator of a lower hydrogen number density used in the heliospheric model than it is in reality. The other significant inconsistency compared to the {\it IBEX-Hi} data is the low fluxes simulated in the frame of the <<power-law tail>> scenario at $\sim$0.71 keV energy channel. This difference, as well as systematically lower model fluxes in the whole {\it IBEX-Hi} energy range, may be induced by several assumptions made in our work, such as (a) the neglect of the velocity (energy) diffusion, (b) the isotropic form of the distribution function throughout the heliosphere, and (c) the weak scattering at the termination shock. A detailed study of these discrepancies will be held in future works.
		
	
	
\end{enumerate}


Important to note that the solution presented in the paper is limited to the IBEX-Hi energy range. At higher energies, i.e. 5.2 -- 55 keV \citep[Ion and Neutral Camera (INCA) on the Cassini spacecraft;][]{krimigis2009}, the ENA spectra is much softer \citep{dialynas2013, dialynas2020}. Even though the {\it IBEX-Hi} (0.3 -- 6 keV) covers a significant part of the proton energy spectrum that exhibits the heliosheath properties (such as plasma velocity, the strength of the TS, the effectiveness of ion acceleration to these energies, etc.), to perform a detailed study of the proton acceleration processes, the modelling results should be compared to a combined energy spectrum extended both to the lower \citep[{\it IBEX-Lo}, 0.01 -- 2 keV;][]{fuselier2009} and higher ({\it INCA/Cassini}) energies, but this work is beyond the scope of the paper. To extend the simulations to IBEX-Lo energies, the fluxes of ENAs originated from core SW protons should be calculated. We have not taken these fluxes into account in the present paper since the ENAs from SW protons produce $\sim$0.1 keV fluxes, which is well below the considered energy range. Nevertheless, our model allows simulating these fluxes, as was described in the \citet{baliukin2020} paper.

To simulate a more realistic distribution of pickup protons, a self-consistent kinetic-MHD model of the heliosphere that treats pickup protons kinetically and separately from the solar wind protons should be developed \citep{malama2006, chalov2015}. This model must take into account (i) the acceleration of pickup ions due to interaction with fluctuating heliospheric magnetic field, and (ii) the interaction of PUIs with the heliospheric termination shock. For the latter, a local model of the PUI--TS interaction \citep[similar to the model by][]{giacalone2021} should be run in the vicinity of the termination shock (to capture the acceleration of ions properly) and incorporated into the global simulations of the SW/LISM interaction. Therefore, our future investigations will be directed towards the development of such a self-consistent model.





\section*{Acknowledgements}
The authors would like to acknowledge the IBEX team for preparing and  making  ENA  fluxes  data  available, and especially Dr Nathan Schwadron for providing information on IBEX GDF uncertainties.
The work was performed in the frame of the Russian Science Foundation grant 19-12-00383.


\section*{Data availability}
The data underlying this article will be shared on reasonable request to the corresponding author.




\bibliographystyle{mnras}




\appendix

\section{Results of the fitting procedure on each of the steps} \label{app:fitting_steps}

This appendix provides the set of figures that justify the estimations obtained with two different assumptions on the PUI velocity distribution downstream of the TS and on each of the four fitting steps, as described in Section \ref{sec:fitting}. The best-fitting parameters are compiled in Tables \ref{tab:fitting_powerlaw} and \ref{tab:fitting_bimaxwl}.

\subsection{<<Power-law tail>> scenario}\label{app:powertail}

\begin{figure*}
\includegraphics[width=\textwidth]{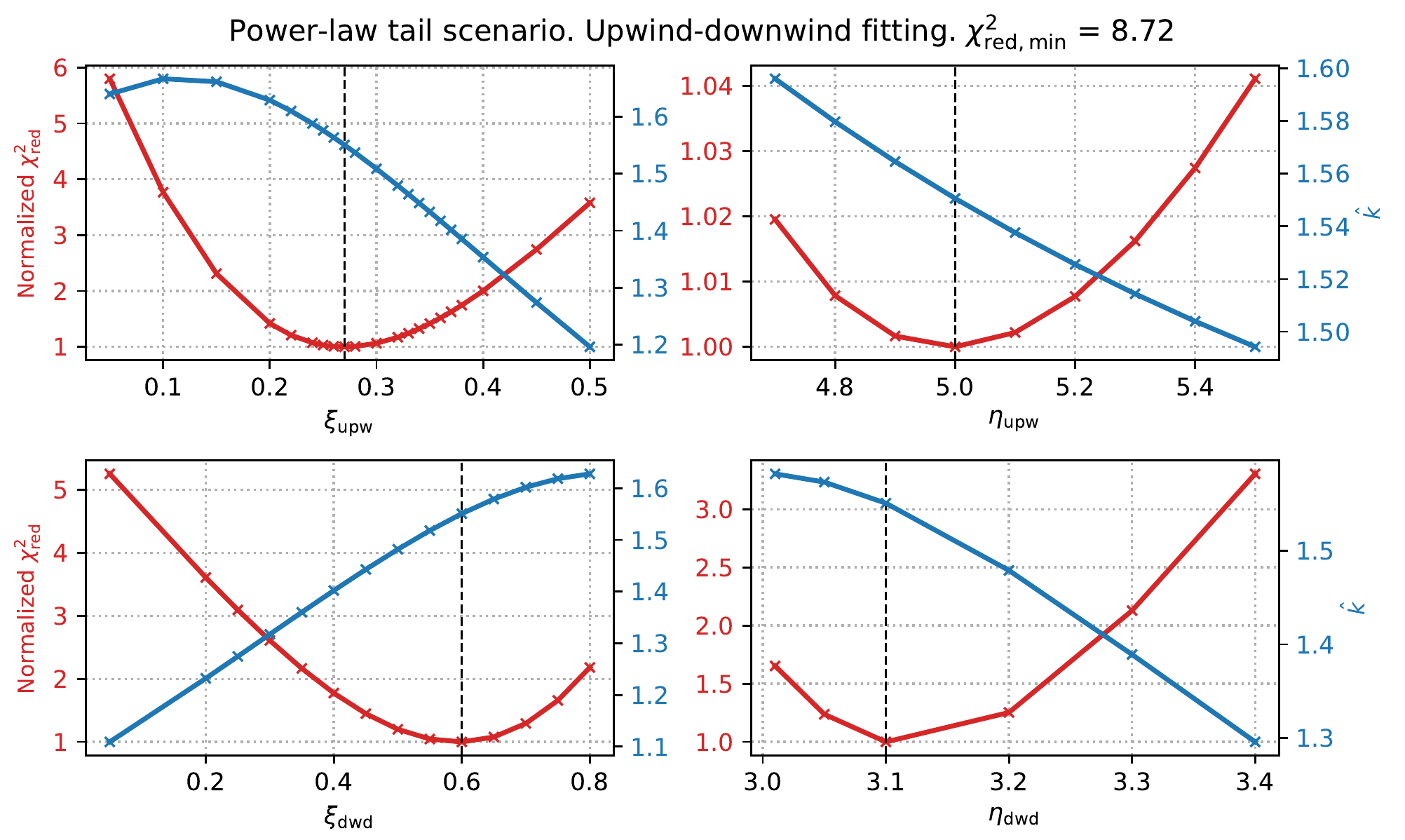}
\caption{The results of the fitting procedure using the <<power-law tail>> scenario for the PUI velocity distribution downstream of the TS on Step 1. In the first and second rows, the dependencies of the normalized $\chi^2_{\rm red}$ statistic (red lines with crosses) and best-fitting scaling coefficient $\hat k$ (blue lines with crosses) on upwind ($\xi_{\rm upw}$, $\eta_{\rm upw}$) and downwind ($\xi_{\rm dwd}$, $\eta_{\rm dwd}$) parameters, respectively, are presented. The best-fitting values ($\hat \xi_{\rm upw} = 0.27$, $\hat \eta_{\rm upw} = 5.0$, $\hat \xi_{\rm dwd} = 0.6$, $\hat \eta_{\rm dwd} = 3.1$), for which the best-fitting scaling coefficient $\hat k = 1.55$, are shown with the black dashed lines.
}
\label{fig:powtail_step1}
\end{figure*}

Figure \ref{fig:powtail_step1} presents the results of the fitting procedure using the <<power-law tail>> scenario for the PUI velocity distribution downstream of the TS on Step 1. In the first and second rows, the dependencies of the normalized $\chi^2_{\rm red}$ statistic (red lines with crosses) and best-fitting scaling coefficient $\hat k$ (blue lines with crosses) on upwind ($\xi_{\rm upw}$, $\eta_{\rm upw}$) and downwind ($\xi_{\rm dwd}$, $\eta_{\rm dwd}$) parameters, respectively, are presented. The $\chi^2_{\rm red}$ was normalized by its minimal value $\chi^2_{\rm red,min} = 8.72$. The best-fitting values ($\hat \xi_{\rm upw} = 0.27$, $\hat \eta_{\rm upw} = 5.0$, $\hat \xi_{\rm dwd} = 0.6$, $\hat \eta_{\rm dwd} = 3.1$), for which the best-fitting scaling coefficient $\hat k = 1.55$, are shown with the black dashed lines.

\begin{figure*}
\includegraphics[width=\textwidth]{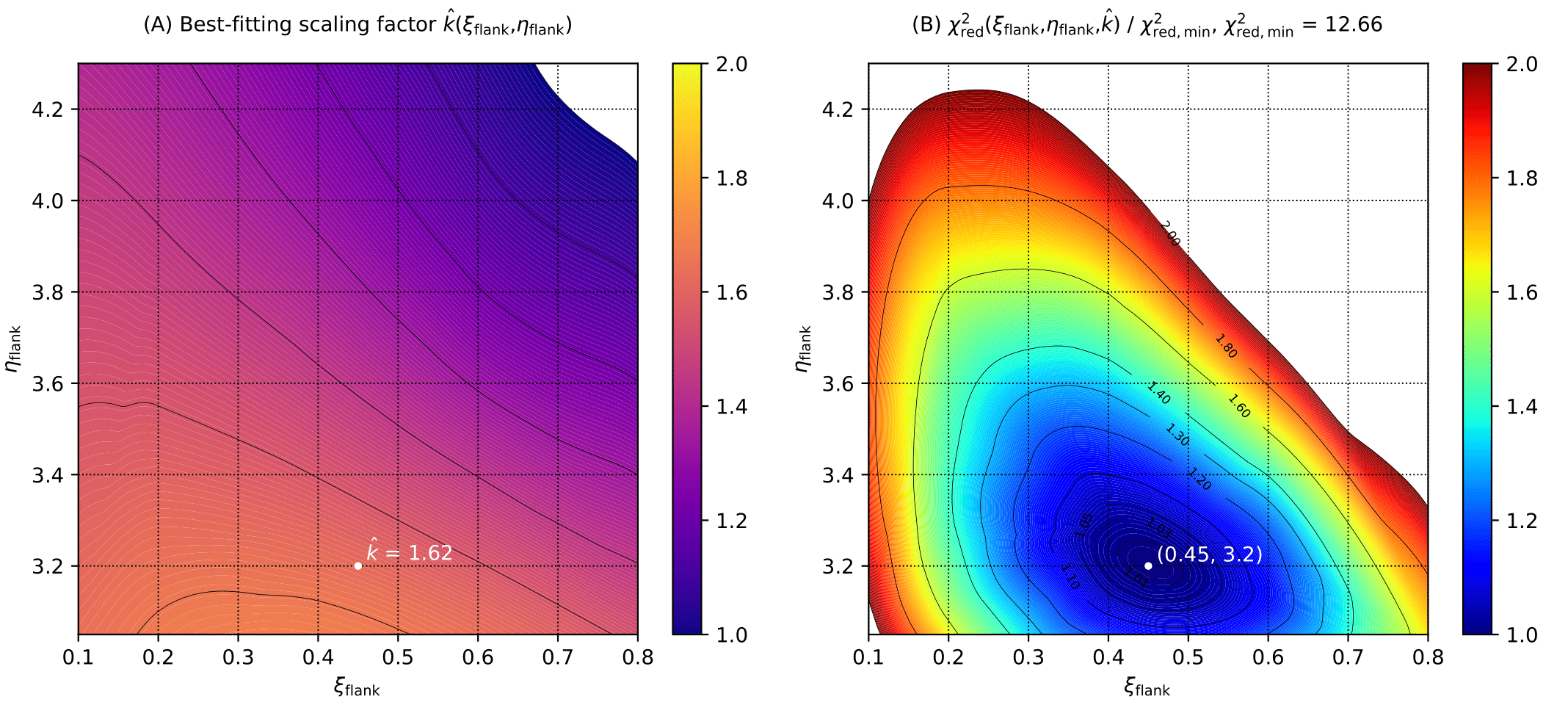}
\caption{The results of the fitting procedure using the <<power-law tail>> scenario for the PUI velocity distribution downstream of the TS on Step 2. The best-fitting scaling factor $\hat k$ (plot A) and normalized $\chi^2_{\rm red}$ statistic (plot B) are presented depending on parameters in the flanks ($\xi_{\rm flank}$, $\eta_{\rm flank}$). The best-fitting values ($\hat \xi_{\rm flank} = 0.45$, $\hat \eta_{\rm flank} = 3.2$), for which the best-fitting scaling coefficient $\hat k = 1.62$, are shown with white dots.
}
\label{fig:powtail_step2}
\end{figure*}

In Figure \ref{fig:powtail_step2} the results of Step 2 are shown. The best-fitting scaling factor $\hat k$ (plot A) and normalized $\chi^2_{\rm red}$ statistic (plot B) are presented depending on parameters in the flanks ($\xi_{\rm flank}$, $\eta_{\rm flank}$). The $\chi^2_{\rm red}$ was normalized by its minimal value $\chi^2_{\rm red,min} = 12.66$. The best-fitting values ($\hat \xi_{\rm flank} = 0.45$, $\hat \eta_{\rm flank} = 3.2$), for which the best-fitting scaling coefficient $\hat k = 1.62$, are shown with white dots.

\begin{figure*}
\includegraphics[width=\textwidth]{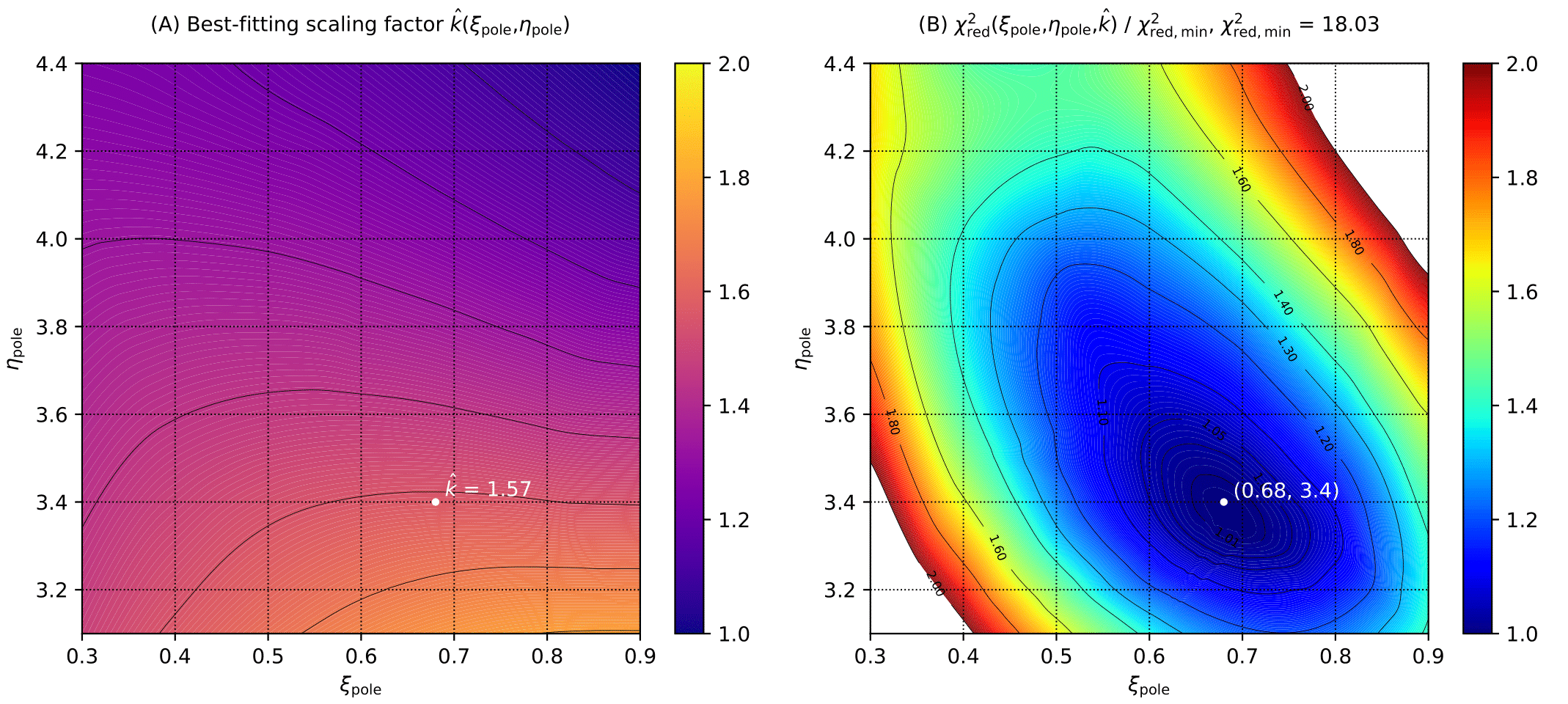}
\caption{The results of the fitting procedure using the <<power-law tail>> scenario for the PUI velocity distribution downstream of the TS on Step 3. The best-fitting scaling factor $\hat k$ (plot A) and normalized $\chi^2_{\rm red}$ statistic (plot B) are presented depending on parameters in the poles ($\xi_{\rm pole}$, $\eta_{\rm pole}$). The best-fitting values ($\hat \xi_{\rm pole} = 0.68$, $\hat \eta_{\rm flank} = 3.4$), for which the best-fitting scaling coefficient $\hat k = 1.57$, are shown with white dots.
}
\label{fig:powtail_step3}
\end{figure*}

Figure \ref{fig:powtail_step3} presents the results of Step 3. The best-fitting scaling factor $\hat k$ (plot A) and normalized $\chi^2_{\rm red}$ statistic (plot B) are presented depending on parameters in the poles ($\xi_{\rm pole}$, $\eta_{\rm pole}$). The $\chi^2_{\rm red}$ was normalized by its minimal value $\chi^2_{\rm red,min} = 18.03$. The best-fitting values ($\hat \xi_{\rm pole} = 0.68$, $\hat \eta_{\rm flank} = 3.4$), for which the best-fitting scaling coefficient $\hat k = 1.57$, are shown with white dots.

\begin{figure*}
\includegraphics[width=\textwidth]{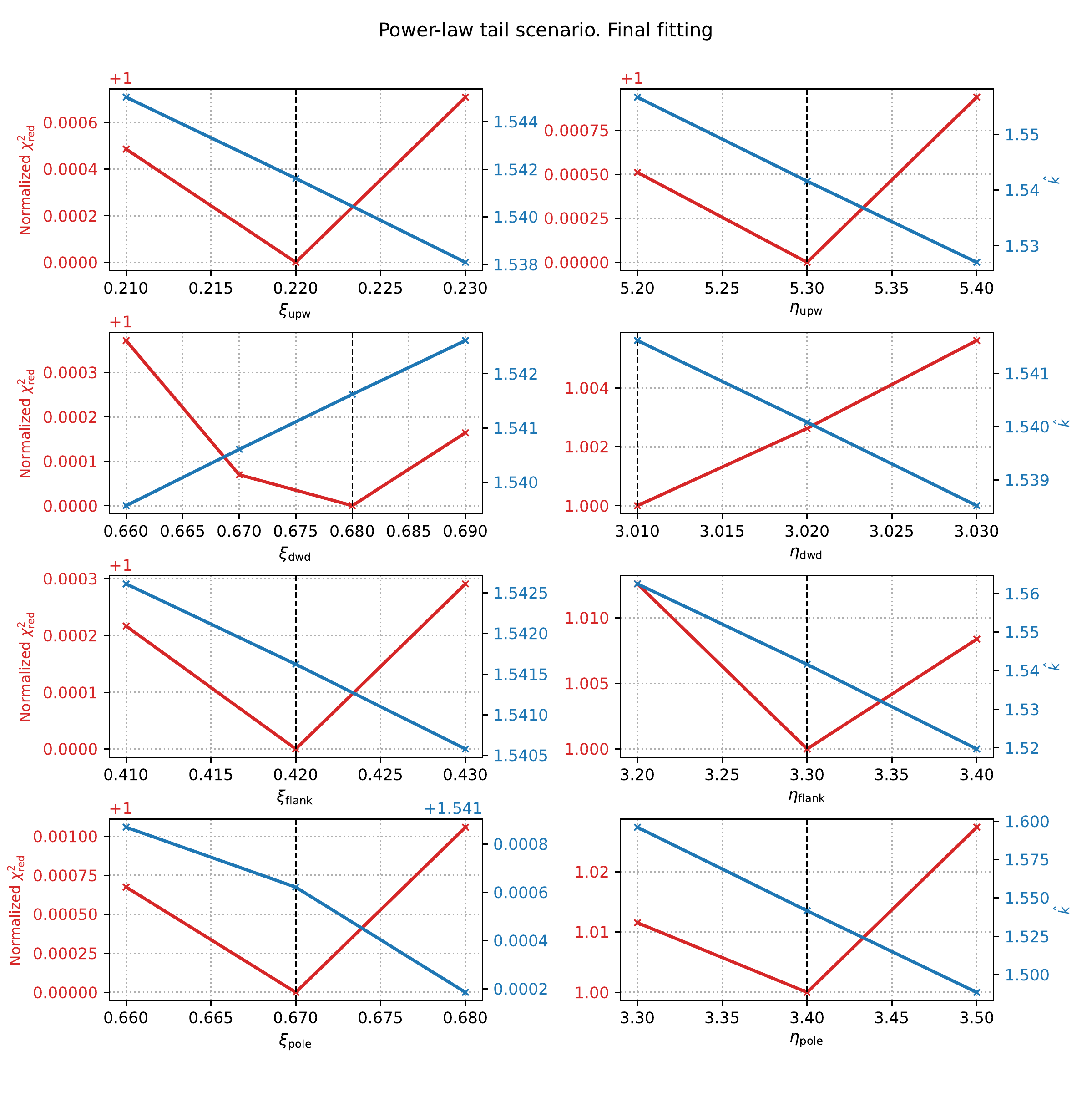}
\caption{The results of the fitting procedure using the <<power-law tail>> scenario for the PUI velocity distribution downstream of the TS on Step 4. In the first, second, third, and fourth rows the dependencies of the normalized $\chi^2_{\rm red}$ statistic (red lines with crosses) and best-fitting scaling coefficient $\hat k$ (blue lines with crosses) on upwind ($\xi_{\rm upw}$, $\eta_{\rm upw}$), downwind ($\xi_{\rm dwd}$, $\eta_{\rm dwd}$), flank ($\xi_{\rm flank}$, $\eta_{\rm flank}$), and pole ($\xi_{\rm pole}$, $\eta_{\rm pole}$) parameters, respectively, are presented. The corresponding $\xi$ and $\eta$ parameters are shown in the first and second columns, respectively. The best-fitting values ($\hat \xi_{\rm upw} = 0.22$, $\hat \eta_{\rm upw} = 5.3$, $\hat \xi_{\rm dwd} = 0.68$, $\hat \eta_{\rm dwd} = 3.01$, $\hat \xi_{\rm flank} = 0.42$, $\hat \eta_{\rm flank} = 3.3$, $\hat \xi_{\rm pole} = 0.67$, $\hat \eta_{\rm pole} = 3.4$), for which the best-fitting scaling coefficient $\hat k = 1.54$, are shown with the black dashed lines.
}
\label{fig:powtail_stepf}
\end{figure*}

The results of the final Step 4 are presented in Figure \ref{fig:powtail_stepf}. In the first, second, third, and fourth rows the dependencies of the normalized $\chi^2_{\rm red}$ statistic (red lines with crosses) and best-fitting scaling coefficient $\hat k$ (blue lines with crosses) on upwind ($\xi_{\rm upw}$, $\eta_{\rm upw}$), downwind ($\xi_{\rm dwd}$, $\eta_{\rm dwd}$), flank ($\xi_{\rm flank}$, $\eta_{\rm flank}$), and pole ($\xi_{\rm pole}$, $\eta_{\rm pole}$) parameters, respectively, are presented. The corresponding $\xi$ and $\eta$ parameters are shown in the first and second columns, respectively. The $\chi^2_{\rm red}$ was normalized by its minimal value $\chi^2_{\rm red,min} = 17.05$. The best-fitting values ($\hat \xi_{\rm upw} = 0.22$, $\hat \eta_{\rm upw} = 5.3$, $\hat \xi_{\rm dwd} = 0.68$, $\hat \eta_{\rm dwd} = 3.01$, $\hat \xi_{\rm flank} = 0.42$, $\hat \eta_{\rm flank} = 3.3$, $\hat \xi_{\rm pole} = 0.67$, $\hat \eta_{\rm pole} = 3.4$), for which the best-fitting scaling coefficient $\hat k = 1.54$, are shown with the black dashed lines.

\subsection{<<Bi-Maxwellian>> scenario}\label{app:bi-maxwl}

\begin{figure*}
\includegraphics[width=\textwidth]{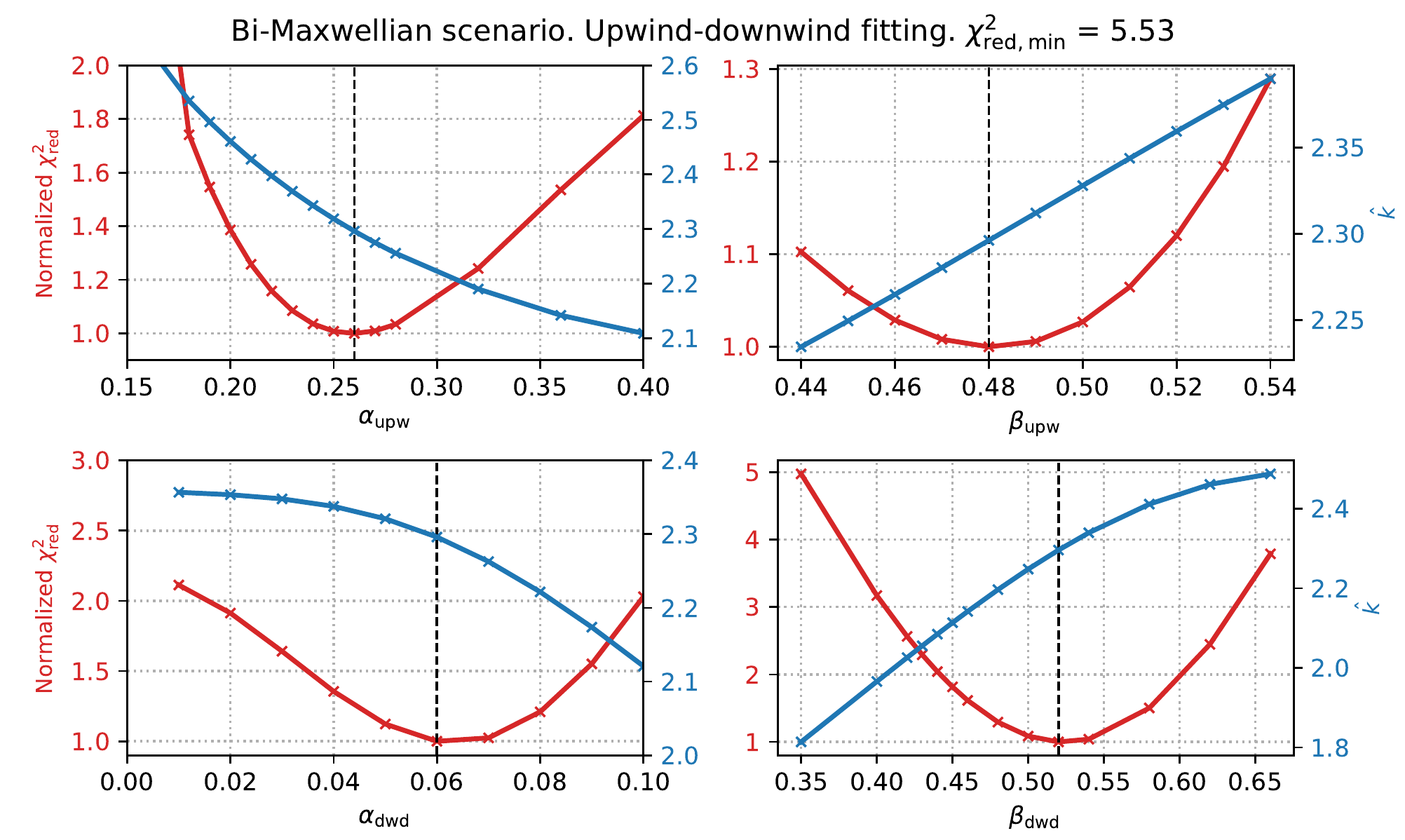}
\caption{The results of the fitting procedure using the <<bi-Maxwellian>> scenario for the PUI velocity distribution downstream of the TS on Step 1. In the first and second rows the dependencies of the $\chi^2_{\rm red}$ statistic (red lines with crosses) and best-fitting scaling coefficient $\hat k$ (blue lines with crosses) on upwind ($\alpha_{\rm upw}$, $\beta_{\rm upw}$) and downwind ($\alpha_{\rm dwd}$, $\beta_{\rm dwd}$) parameters, respectively, are presented. The best-fitting values ($\hat \alpha_{\rm upw} = 0.26$, $\hat \beta_{\rm upw} = 0.48$, $\hat \alpha_{\rm dwd} = 0.06$, $\hat \beta_{\rm dwd} = 0.52$), for which the best-fitting scaling coefficient $\hat k = 2.3$, are shown with the black dashed lines.
}
\label{fig:bimaxwl_step1}
\end{figure*}

Figure \ref{fig:bimaxwl_step1} presents the results of the fitting procedure using the <<bi-Maxwellian>> scenario for the PUI velocity distribution downstream of the TS on Step 1. In the first and second rows, the dependencies of the $\chi^2_{\rm red}$ statistic (red lines with crosses) and best-fitting scaling coefficient $\hat k$ (blue lines with crosses) on upwind ($\alpha_{\rm upw}$, $\beta_{\rm upw}$) and downwind ($\alpha_{\rm dwd}$, $\beta_{\rm dwd}$) parameters, respectively, are presented. The $\chi^2_{\rm red}$ was normalized by its minimal value $\chi^2_{\rm red,min} = 5.53$. The best-fitting values ($\hat \alpha_{\rm upw} = 0.26$, $\hat \beta_{\rm upw} = 0.48$, $\hat \alpha_{\rm dwd} = 0.06$, $\hat \beta_{\rm dwd} = 0.52$), for which the best-fitting scaling coefficient $\hat k = 2.3$, are shown with the black dashed lines.

\begin{figure*}
\includegraphics[width=\textwidth]{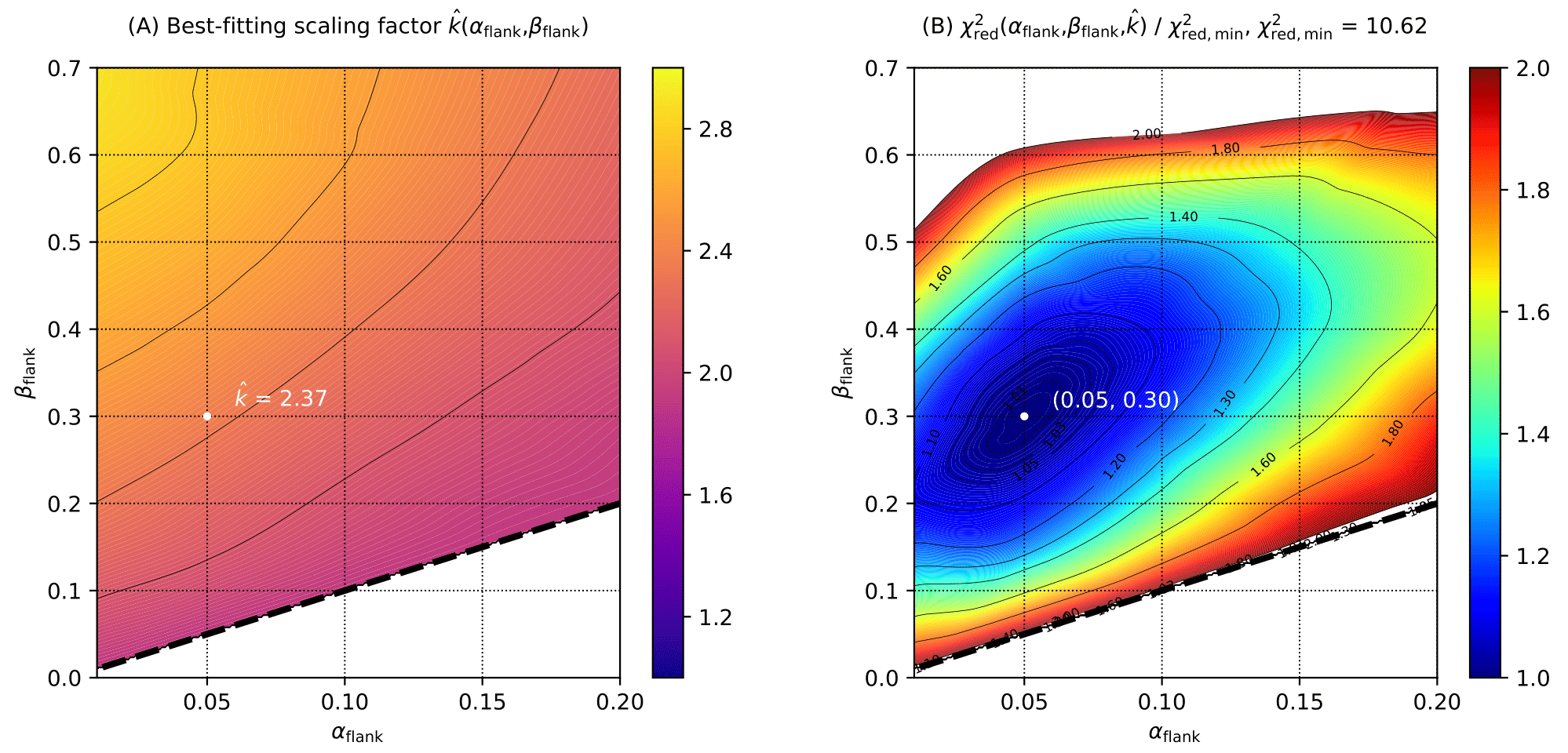}
\caption{The results of the fitting procedure using the <<bi-Maxwellian>> scenario for the PUI velocity distribution downstream of the TS on Step 2. The best-fitting scaling factor $\hat k$ (plot A) and normalized $\chi^2_{\rm red}$ statistic (plot B) are presented depending on parameters in the flanks ($\alpha_{\rm flank}$, $\beta_{\rm flank}$). The best-fitting values ($\hat \alpha_{\rm flank} = 0.05$, $\hat \beta_{\rm flank} = 0.3$), for which the best-fitting scaling coefficient $\hat k = 2.37$, are shown with white dots.
}
\label{fig:bimaxwl_step2}
\end{figure*}

In Figure \ref{fig:bimaxwl_step2} the results of Step 2 are shown. The best-fitting scaling factor $\hat k$ (plot A) and normalized $\chi^2_{\rm red}$ statistic (plot B) are presented depending on parameters in the flanks ($\alpha_{\rm flank}$, $\beta_{\rm flank}$). The $\chi^2_{\rm red}$ was normalized by its minimal value $\chi^2_{\rm red,min} = 10.62$. The best-fitting values ($\hat \alpha_{\rm flank} = 0.05$, $\hat \beta_{\rm flank} = 0.3$), for which the best-fitting scaling coefficient $\hat k = 2.37$, are shown with white dots.

\begin{figure*}
\includegraphics[width=\textwidth]{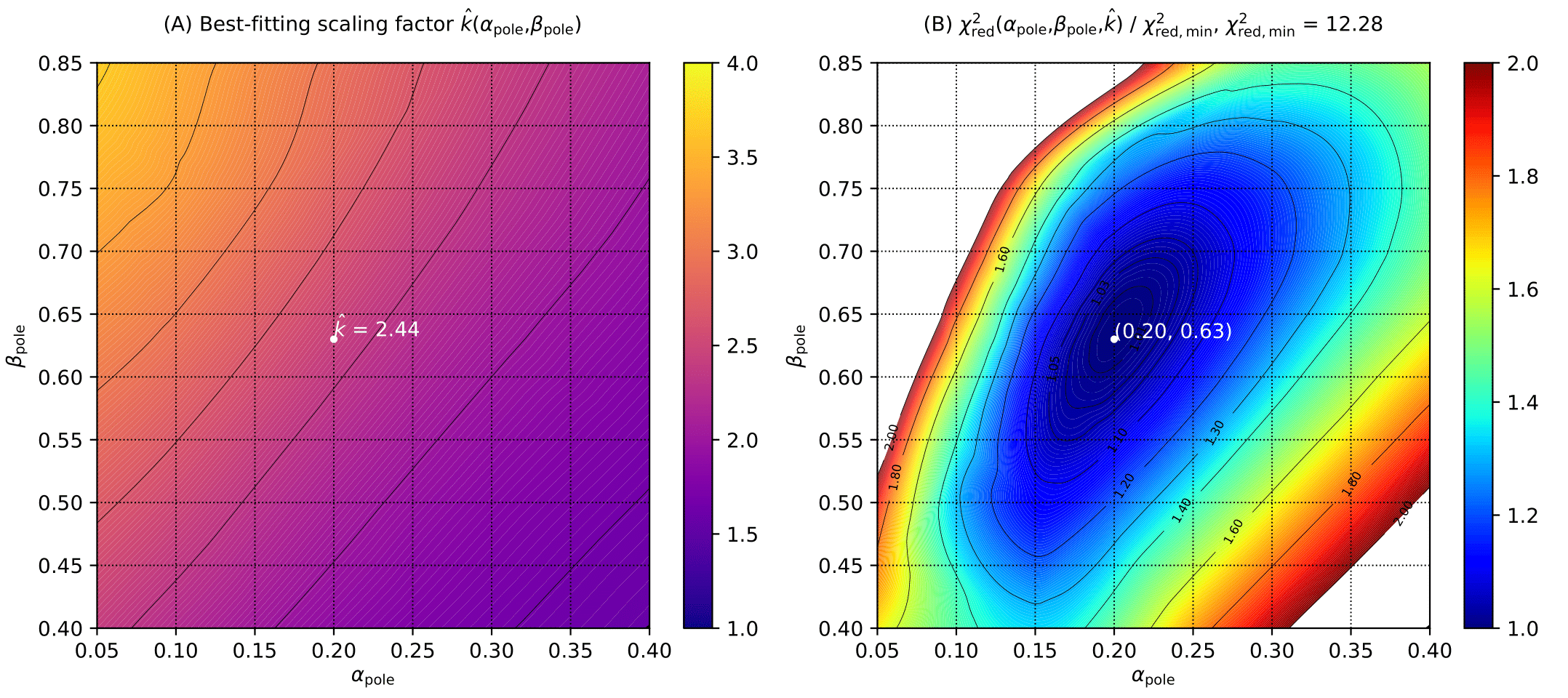}
\caption{The results of the fitting procedure using the <<bi-Maxwellian>> scenario for the PUI velocity distribution downstream of the TS on Step 3. The best-fitting scaling factor $\hat k$ (plot A) and normalized $\chi^2_{\rm red}$ statistic (plot B) are presented depending on parameters in the poles ($\alpha_{\rm pole}$, $\beta_{\rm pole}$). The best-fitting values ($\hat \alpha_{\rm pole} = 0.2$, $\hat \beta_{\rm flank} = 0.63$), for which the best-fitting scaling coefficient $\hat k = 2.44$, are shown with white dots.
}
\label{fig:bimaxwl_step3}
\end{figure*}

Figure \ref{fig:bimaxwl_step3} presents the results of Step 3. The best-fitting scaling factor $\hat k$ (plot A) and normalized $\chi^2_{\rm red}$ statistic (plot B) are presented depending on parameters in the poles ($\alpha_{\rm pole}$, $\beta_{\rm pole}$). The $\chi^2_{\rm red}$ was normalized by its minimal value $\chi^2_{\rm red,min} = 12.28$. The best-fitting values ($\hat \alpha_{\rm pole} = 0.2$, $\hat \beta_{\rm flank} = 0.63$), for which the best-fitting scaling coefficient $\hat k = 2.44$, are shown with white dots.

\begin{figure*}
\includegraphics[width=\textwidth]{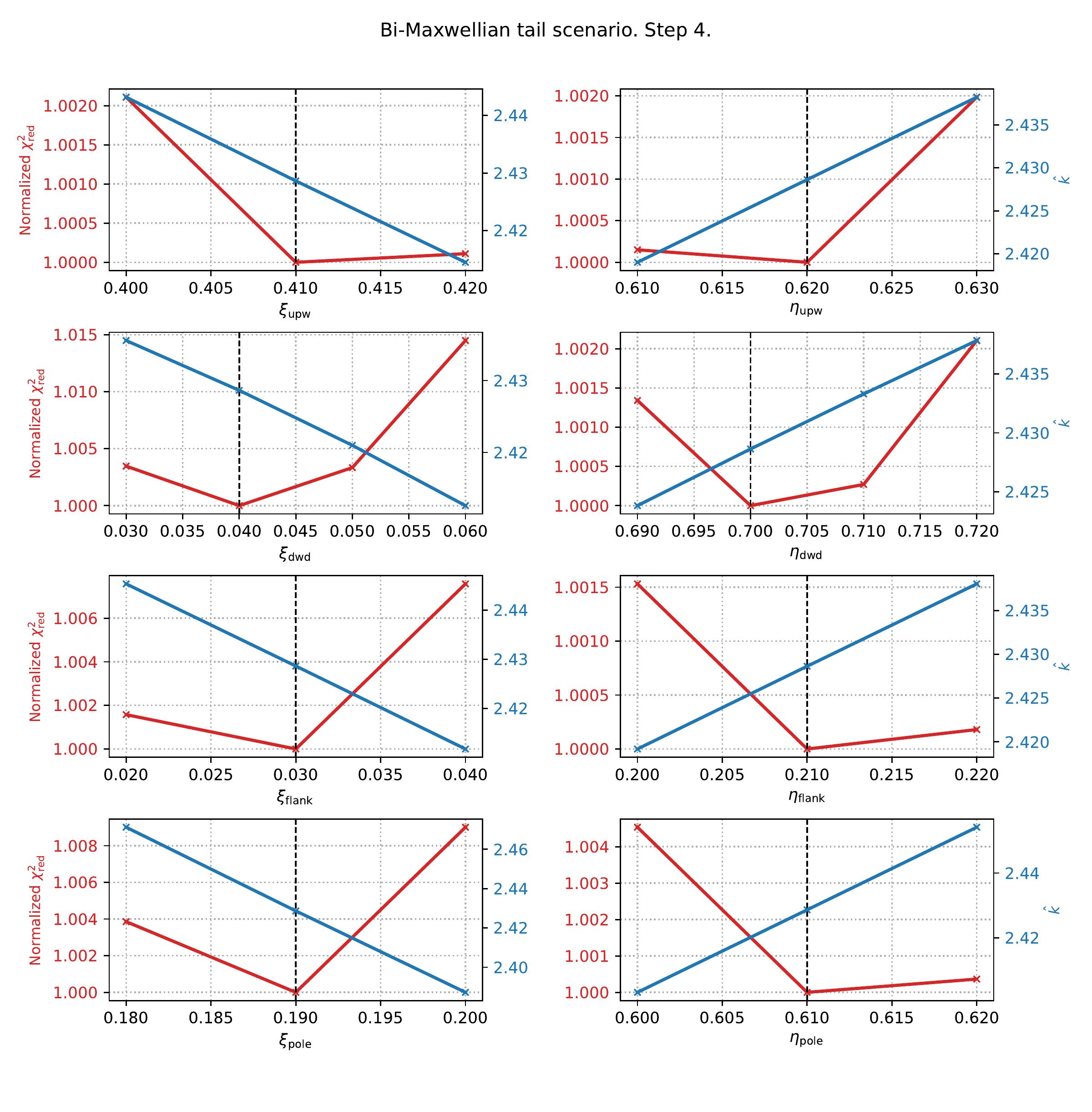}
\caption{The results of the fitting procedure using the <<bi-Maxwellian>> scenario for the PUI velocity distribution downstream of the TS on Step 4. In the first, second, third, and fourth rows the dependencies of the normalized $\chi^2_{\rm red}$ statistic (red lines with crosses) and best-fitting scaling coefficient $\hat k$ (blue lines with crosses) on upwind ($\alpha_{\rm upw}$, $\beta_{\rm upw}$), downwind ($\alpha_{\rm dwd}$, $\beta_{\rm dwd}$), flank ($\alpha_{\rm flank}$, $\beta_{\rm flank}$), and pole ($\alpha_{\rm pole}$, $\beta_{\rm pole}$) parameters, respectively, are presented. The corresponding $\alpha$ and $\beta$ parameters are shown in the first and second columns, respectively. The best-fitting values ($\hat \alpha_{\rm upw} = 0.41$, $\hat \beta_{\rm upw} = 0.62$, $\hat \alpha_{\rm dwd} = 0.04$, $\hat \beta_{\rm dwd} = 0.7$, $\hat \alpha_{\rm flank} = 0.03$, $\hat \beta_{\rm flank} = 0.21$, $\hat \alpha_{\rm pole} = 0.19$, $\hat \beta_{\rm pole} = 0.61$), for which the best-fitting scaling coefficient $\hat k = 2.43$, are shown with the black dashed lines.
}
\label{fig:bimaxwl_stepf}
\end{figure*}

The results of the final Step 4 are presented in Figure \ref{fig:bimaxwl_stepf}. In the first, second, third, and fourth rows the dependencies of the normalized $\chi^2_{\rm red}$ statistic (red lines with crosses) and best-fitting scaling coefficient $\hat k$ (blue lines with crosses) on upwind ($\alpha_{\rm upw}$, $\beta_{\rm upw}$), downwind ($\alpha_{\rm dwd}$, $\beta_{\rm dwd}$), flank ($\alpha_{\rm flank}$, $\beta_{\rm flank}$), and pole ($\alpha_{\rm pole}$, $\beta_{\rm pole}$) parameters, respectively, are presented. The corresponding $\alpha$ and $\beta$ parameters are shown in the first and second columns, respectively. The $\chi^2_{\rm red}$ was normalized by its minimal value $\chi^2_{\rm red,min} = 7.77$. The best-fitting values ($\hat \alpha_{\rm upw} = 0.41$, $\hat \beta_{\rm upw} = 0.62$, $\hat \alpha_{\rm dwd} = 0.04$, $\hat \beta_{\rm dwd} = 0.7$, $\hat \alpha_{\rm flank} = 0.03$, $\hat \beta_{\rm flank} = 0.21$, $\hat \alpha_{\rm pole} = 0.19$, $\hat \beta_{\rm pole} = 0.61$), for which the best-fitting scaling coefficient $\hat k = 2.43$, are shown with the black dashed lines.

\bsp	
\label{lastpage}
\end{document}